
\input fontch.tex

%
%
%
\def\unredoffs{} \def\redoffs{\voffset=-.31truein\hoffset=-.48truein}
\def\speclscape{}
%
%
%
%
%
\newbox\leftpage \newdimen\fullhsize \newdimen\hstitle \newdimen\hsbody
\tolerance=1000\hfuzz=2pt
\catcode`\@=11 
\ifx\hyperdef\UNd@FiNeD\def\hyperdef#1#2#3#4{#4}\def\hyperref#1#2#3#4{#4}\fi
\def\bigans{b }
\def\answ{b }
%
\ifx\answ\bigans\message{(This will come out unreduced.}
\magnification=1200\unredoffs\baselineskip=16pt plus 2pt minus 1pt
\hsbody=\hsize \hstitle=\hsize 
\else\message{(This will be reduced.} \let\l@r=L
\magnification=1000\baselineskip=16pt plus 2pt minus 1pt \vsize=7truein
\redoffs \hstitle=8truein\hsbody=4.75truein\fullhsize=10truein\hsize=\hsbody
\output={\ifnum\pageno=0 
  \shipout\vbox{\speclscape{\hsize\fullhsize\makeheadline}
    \hbox to \fullhsize{\hfill\pagebody\hfill}}\advancepageno
  \else
  \almostshipout{\leftline{\vbox{\pagebody\makefootline}}}\advancepageno
  \fi}
\def\almostshipout#1{\if L\l@r \count1=1 \message{[\the\count0.\the\count1]}
      \global\setbox\leftpage=#1 \global\let\l@r=R
 \else \count1=2
  \shipout\vbox{\speclscape{\hsize\fullhsize\makeheadline}
      \hbox to\fullhsize{\box\leftpage\hfil#1}}  \global\let\l@r=L\fi}
\fi
%
\newcount\yearltd\yearltd=\year\advance\yearltd by -1900

\def\Title#1#2{\nopagenumbers\abstractfont\hsize=\hstitle\rightline{#1}%
\vskip 1in\centerline{\titlefont #2}\abstractfont\vskip .5in\pageno=0}
\def\Date#1{\vfill\leftline{#1}\tenpoint\supereject\global\hsize=\hsbody%
\footline={\hss\tenrm\hyperdef\hypernoname{page}\folio\folio\hss}}%
%

\def\draftmode{\message{ DRAFTMODE }\def\draftdate{{\rm preliminary draft:
\number\month/\number\day/\number\yearltd\ \ \hourmin}}%
\headline={\hfil\draftdate}\writelabels\baselineskip=20pt plus 2pt minus 2pt
 {\count255=\time\divide\count255 by 60 \xdef\hourmin{\number\count255}
  \multiply\count255 by-60\advance\count255 by\time
  \xdef\hourmin{\hourmin:\ifnum\count255<10 0\fi\the\count255}}}
\def\nolabels{\def\wrlabeL##1{}\def\eqlabeL##1{}\def\reflabeL##1{}}
\def\writelabels{\def\wrlabeL##1{\leavevmode\vadjust{\rlap{\smash%
{\line{{\escapechar=` \hfill\rlap{\sevenrm\hskip.03in\string##1}}}}}}}%
\def\eqlabeL##1{{\escapechar-1\rlap{\sevenrm\hskip.05in\string##1}}}%
\def\reflabeL##1{\noexpand\llap{\noexpand\sevenrm\string\string\string##1}}}
\nolabels
%
\global\newcount\secno \global\secno=0
\global\newcount\meqno \global\meqno=1
\def\s@csym{}
\def\newsec#1{\global\advance\secno by1%
{\toks0{#1}\message{(\the\secno. \the\toks0)}}%
\global\subsecno=0\eqnres@t\let\s@csym\secsym\xdef\secn@m{\the\secno}\noindent
{\bf\hyperdef\hypernoname{section}{\the\secno}{\the\secno.} #1}%
\writetoca{{\string\hyperref{}{section}{\the\secno}{\the\secno.}} {#1}}%
\par\nobreak\medskip\nobreak}
\def\eqnres@t{\xdef\secsym{\the\secno.}\global\meqno=1\bigbreak\bigskip}
\def\sequentialequations{\def\eqnres@t{\bigbreak}}\xdef\secsym{}
\global\newcount\subsecno \global\subsecno=0
\def\subsec#1{\global\advance\subsecno by1%
{\toks0{#1}\message{(\s@csym\the\subsecno. \the\toks0)}}%
\ifnum\lastpenalty>9000\else\bigbreak\fi
\noindent{\it\hyperdef\hypernoname{subsection}{\secn@m.\the\subsecno}%
{\secn@m.\the\subsecno.} #1}\writetoca{\string\quad
{\string\hyperref{}{subsection}{\secn@m.\the\subsecno}{\secn@m.\the\subsecno.}}
{#1}}\par\nobreak\medskip\nobreak}
\def\appendix#1#2{\global\meqno=1\global\subsecno=0\xdef\secsym{\hbox{#1.}}%
\bigbreak\bigskip\noindent{\bf Appendix \hyperdef\hypernoname{appendix}{#1}%
{#1.} #2}{\toks0{(#1. #2)}\message{\the\toks0}}%
\xdef\s@csym{#1.}\xdef\secn@m{#1}%
\writetoca{\string\hyperref{}{appendix}{#1}{Appendix {#1.}} {#2}}%
\par\nobreak\medskip\nobreak}
%
%
\def\checkm@de#1#2{\ifmmode{\def\f@rst##1{##1}\hyperdef\hypernoname{equation}%
{#1}{#2}}\else\hyperref{}{equation}{#1}{#2}\fi}
\def\eqnn#1{\DefWarn#1\xdef #1{(\noexpand\relax\noexpand\checkm@de%
{\s@csym\the\meqno}{\secsym\the\meqno})}%
\wrlabeL#1\writedef{#1\leftbracket#1}\global\advance\meqno by1}
\def\f@rst#1{\c@t#1a\em@ark}\def\c@t#1#2\em@ark{#1}
\def\eqna#1{\DefWarn#1\wrlabeL{#1$\{\}$}%
\xdef #1##1{(\noexpand\relax\noexpand\checkm@de%
{\s@csym\the\meqno\noexpand\f@rst{##1}}{\hbox{$\secsym\the\meqno##1$}})}
\writedef{#1\numbersign1\leftbracket#1{\numbersign1}}\global\advance\meqno by1}
\def\eqn#1#2{\DefWarn#1%
\xdef #1{(\noexpand\hyperref{}{equation}{\s@csym\the\meqno}%
{\secsym\the\meqno})}$$#2\eqno(\hyperdef\hypernoname{equation}%
{\s@csym\the\meqno}{\secsym\the\meqno})\eqlabeL#1$$%
\writedef{#1\leftbracket#1}\global\advance\meqno by1}
\def\xeqn{\expandafter\xe@n}\def\xe@n(#1){#1}
\def\xeqna#1{\expandafter\xe@n#1}
\def\eqns#1{(\e@ns #1{\hbox{}})}
\def\e@ns#1{\ifx\UNd@FiNeD#1\message{eqnlabel \string#1 is undefined.}%
\xdef#1{(?.?)}\fi{\let\hyperref=\relax\xdef\next{#1}}%
\ifx\next\em@rk\def\next{}\else%
\ifx\next#1\xeqn#1\else\def\n@xt{#1}\ifx\n@xt\next#1\else\xeqna#1\fi
\fi\let\next=\e@ns\fi\next}

\def\DefWarn#1{\ifx\UNd@FiNeD#1\else
\immediate\write16{*** WARNING: the label \string#1 is already defined ***}\fi}
%
\newskip\footskip\footskip14pt plus 1pt minus 1pt 
\def\footnotefont{\ninepoint}\def\f@t#1{\footnotefont #1\@foot}
\def\f@@t{\baselineskip\footskip\bgroup\footnotefont\aftergroup\@foot\let\next}
\setbox\strutbox=\hbox{\vrule height9.5pt depth4.5pt width0pt}
\global\newcount\ftno \global\ftno=0
\def\foot{\global\advance\ftno by1\def\foot@rg{\hyperref{}{footnote}%
{\the\ftno}{\the\ftno}\xdef\foot@rg{\noexpand\hyperdef\noexpand\hypernoname%
{footnote}{\the\ftno}{\the\ftno}}}\footnote{$^{\foot@rg}$}}
%
\newwrite\ftfile
\def\footend{\def\foot{\global\advance\ftno by1\chardef\wfile=\ftfile
\hyperref{}{footnote}{\the\ftno}{$^{\the\ftno}$}%
\ifnum\ftno=1\immediate\openout\ftfile=\jobname.fts\fi%
\immediate\write\ftfile{\noexpand\smallskip%
\noexpand\item{\noexpand\hyperdef\noexpand\hypernoname{footnote}
{\the\ftno}{f\the\ftno}:\ }\pctsign}\findarg}%
\def\footatend{\vfill\eject\immediate\closeout\ftfile{\parindent=20pt
\centerline{\bf Footnotes}\nobreak\bigskip\input \jobname.fts }}}
\def\footatend{}
%
%
\global\newcount\refno \global\refno=1
\newwrite\rfile
\def\ref{[\hyperref{}{reference}{\the\refno}{\the\refno}]\nref}
\def\nref#1{\DefWarn#1%
\xdef#1{[\noexpand\hyperref{}{reference}{\the\refno}{\the\refno}]}%
\writedef{#1\leftbracket#1}%
\ifnum\refno=1\immediate\openout\rfile=\jobname.refs\fi
\chardef\wfile=\rfile\immediate\write\rfile{\noexpand\item{[\noexpand\hyperdef%
\noexpand\hypernoname{reference}{\the\refno}{\the\refno}]\ }%
\reflabeL{#1\hskip.31in}\pctsign}\global\advance\refno by1\findarg}
\def\findarg#1#{\begingroup\obeylines\newlinechar=`\^^M\pass@rg}
{\obeylines\gdef\pass@rg#1{\writ@line\relax #1^^M\hbox{}^^M}%
\gdef\writ@line#1^^M{\expandafter\toks0\expandafter{\striprel@x #1}%
\edef\next{\the\toks0}\ifx\next\em@rk\let\next=\endgroup\else\ifx\next\empty%
\else\immediate\write\wfile{\the\toks0}\fi\let\next=\writ@line\fi\next\relax}}
\def\striprel@x#1{} \def\em@rk{\hbox{}}
\def\lref{\begingroup\obeylines\lr@f}
\def\lr@f#1#2{\DefWarn#1\gdef#1{\let#1=\UNd@FiNeD\ref#1{#2}}\endgroup\unskip}

\def\addref#1{\immediate\write\rfile{\noexpand\item{}#1}} 
\def\listrefs{\footatend\vfill\supereject\immediate\closeout\rfile\writestoppt
\baselineskip=\footskip\centerline{{\bf References}}\bigskip{\parindent=20pt%
\frenchspacing\escapechar=` \input \jobname.refs\vfill\eject}\nonfrenchspacing}
\def\startrefs#1{\immediate\openout\rfile=\jobname.refs\refno=#1}
\def\xref{\expandafter\xr@f}\def\xr@f[#1]{#1}
\def\refs#1{\count255=1[\r@fs #1{\hbox{}}]}
\def\r@fs#1{\ifx\UNd@FiNeD#1\message{reflabel \string#1 is undefined.}%
\nref#1{need to supply reference \string#1.}\fi%
\vphantom{\hphantom{#1}}{\let\hyperref=\relax\xdef\next{#1}}%
\ifx\next\em@rk\def\next{}%
\else\ifx\next#1\ifodd\count255\relax\xref#1\count255=0\fi%
\else#1\count255=1\fi\let\next=\r@fs\fi\next}
%

%
\newwrite\ffile\global\newcount\figno \global\figno=1
\def\fig{fig.~\hyperref{}{figure}{\the\figno}{\the\figno}\nfig}
\def\nfig#1{\DefWarn#1%
\xdef#1{fig.~\noexpand\hyperref{}{figure}{\the\figno}{\the\figno}}%
\writedef{#1\leftbracket fig.\noexpand~\xfig#1}%
\ifnum\figno=1\immediate\openout\ffile=\jobname.figs\fi\chardef\wfile=\ffile%
{\let\hyperref=\relax
\immediate\write\ffile{\noexpand\medskip\noexpand\item{Fig.\ %
\noexpand\hyperdef\noexpand\hypernoname{figure}{\the\figno}{\the\figno}. }
\reflabeL{#1\hskip.55in}\pctsign}}\global\advance\figno by1\findarg}
\def\listfigs{\vfill\eject\immediate\closeout\ffile{\parindent40pt
\baselineskip14pt\centerline{{\bf Figure Captions}}\nobreak\medskip
\escapechar=` \input \jobname.figs\vfill\eject}}
\def\xfig{\expandafter\xf@g}\def\xf@g fig.\penalty\@M\ {}
\def\figs#1{figs.~\f@gs #1{\hbox{}}}
\def\f@gs#1{{\let\hyperref=\relax\xdef\next{#1}}\ifx\next\em@rk\def\next{}\else
\ifx\next#1\xfig #1\else#1\fi\let\next=\f@gs\fi\next}
\def\figin{\epsfcheck\figin}\def\figins{\epsfcheck\figins}
\def\epsfcheck{\ifx\epsfbox\UNd@FiNeD
\message{(NO epsf.tex, FIGURES WILL BE IGNORED)}
\gdef\figin##1{\vskip2in}\gdef\figins##1{\hskip.5in}
\else\message{(FIGURES WILL BE INCLUDED)}%
\gdef\figin##1{##1}\gdef\figins##1{##1}\fi}
\def\DefWarn#1{}
\def\figinsert{\goodbreak\midinsert}
\def\ifig#1#2#3{\DefWarn#1\xdef#1{fig.~\noexpand\hyperref{}{figure}%
{\the\figno}{\the\figno}}\writedef{#1\leftbracket fig.\noexpand~\xfig#1}%
\figinsert\figin{\centerline{#3}}\medskip\centerline{\vbox{\baselineskip12pt
\advance\hsize by -1truein\noindent\wrlabeL{#1=#1}\footnotefont%
{\bf Fig.~\hyperdef\hypernoname{figure}{\the\figno}{\the\figno}:} #2}}
\bigskip\endinsert\global\advance\figno by1}
\newwrite\lfile
{\escapechar-1\xdef\pctsign{\string\%}\xdef\leftbracket{\string\{}
\xdef\rightbracket{\string\}}\xdef\numbersign{\string\#}}
\def\writedefs{\immediate\openout\lfile=\jobname.defs \def\writedef##1{%
{\let\hyperref=\relax\let\hyperdef=\relax\let\hypernoname=\relax
 \immediate\write\lfile{\string\def\string##1\rightbracket}}}}%
\def\writestop{\def\writestoppt{\immediate\write\lfile{\string\pageno
 \the\pageno\string\startrefs\leftbracket\the\refno\rightbracket
 \string\def\string\secsym\leftbracket\secsym\rightbracket
 \string\secno\the\secno\string\meqno\the\meqno}\immediate\closeout\lfile}}
\def\writestoppt{}\def\writedef#1{}
\def\seclab#1{\DefWarn#1%
\xdef #1{\noexpand\hyperref{}{section}{\the\secno}{\the\secno}}%
\writedef{#1\leftbracket#1}\wrlabeL{#1=#1}}
\def\subseclab#1{\DefWarn#1%
\xdef #1{\noexpand\hyperref{}{subsection}{\secn@m.\the\subsecno}%
{\secn@m.\the\subsecno}}\writedef{#1\leftbracket#1}\wrlabeL{#1=#1}}
\def\applab#1{\DefWarn#1%
\xdef #1{\noexpand\hyperref{}{appendix}{\secn@m}{\secn@m}}%
\writedef{#1\leftbracket#1}\wrlabeL{#1=#1}}
\newwrite\tfile \def\writetoca#1{}
\def\leaderfill{\leaders\hbox to 1em{\hss.\hss}\hfill}
\def\writetoc{\immediate\openout\tfile=\jobname.toc
   \def\writetoca##1{{\edef\next{\write\tfile{\noindent ##1
   \string\leaderfill {\string\hyperref{}{page}{\noexpand\number\pageno}%
                       {\noexpand\number\pageno}} \par}}\next}}}
\newread\ch@ckfile
\def\listtoc{\immediate\closeout\tfile\immediate\openin\ch@ckfile=\jobname.toc
\ifeof\ch@ckfile\message{no file \jobname.toc, no table of contents this pass}%
\else\closein\ch@ckfile\centerline{\bf Contents}\nobreak\medskip%
{\baselineskip=12pt\footnotefont\parskip=0pt\catcode`\@=11\input\jobname.toc
\catcode`\@=12\bigbreak\bigskip}\fi}
\catcode`\@=12 
%
\edef\tfontsize{\ifx\answ\bigans scaled\magstep3\else scaled\magstep4\fi}
\font\titlerm=cmr10 \tfontsize \font\titlerms=cmr7 \tfontsize
\font\titlermss=cmr5 \tfontsize \font\titlei=cmmi10 \tfontsize
\font\titleis=cmmi7 \tfontsize \font\titleiss=cmmi5 \tfontsize
\font\titlesy=cmsy10 \tfontsize \font\titlesys=cmsy7 \tfontsize
\font\titlesyss=cmsy5 \tfontsize \font\titleit=cmti10 \tfontsize
\skewchar\titlei='177 \skewchar\titleis='177 \skewchar\titleiss='177
\skewchar\titlesy='60 \skewchar\titlesys='60 \skewchar\titlesyss='60
\def\titlefont{\def\rm{\fam0\titlerm}
\textfont0=\titlerm \scriptfont0=\titlerms \scriptscriptfont0=\titlermss
\textfont1=\titlei \scriptfont1=\titleis \scriptscriptfont1=\titleiss
\textfont2=\titlesy \scriptfont2=\titlesys \scriptscriptfont2=\titlesyss
\textfont\itfam=\titleit \def\it{\fam\itfam\titleit}\rm}
 \ifx\answ\bigans\else scaled\magstep1\fi
\ifx\answ\bigans\def\abstractfont{\tenpoint}\else
\font\absit=cmti10 scaled \magstep1
\font\abssl=cmsl10 scaled \magstep1
\font\absrm=cmr10 scaled\magstep1 \font\absrms=cmr7 scaled\magstep1
\font\absrmss=cmr5 scaled\magstep1 \font\absi=cmmi10 scaled\magstep1
\font\absis=cmmi7 scaled\magstep1 \font\absiss=cmmi5 scaled\magstep1
\font\abssy=cmsy10 scaled\magstep1 \font\abssys=cmsy7 scaled\magstep1
\font\abssyss=cmsy5 scaled\magstep1 \font\absbf=cmbx10 scaled\magstep1
\skewchar\absi='177 \skewchar\absis='177 \skewchar\absiss='177
\skewchar\abssy='60 \skewchar\abssys='60 \skewchar\abssyss='60
\def\abstractfont{\def\rm{\fam0\absrm}
\textfont0=\absrm \scriptfont0=\absrms \scriptscriptfont0=\absrmss
\textfont1=\absi \scriptfont1=\absis \scriptscriptfont1=\absiss
\textfont2=\abssy \scriptfont2=\abssys \scriptscriptfont2=\abssyss
\textfont\itfam=\absit \def\it{\fam\itfam\absit}\def\footnotefont{\tenpoint}%
\textfont\slfam=\abssl \def\sl{\fam\slfam\abssl}%
\textfont\bffam=\absbf \def\bf{\fam\bffam\absbf}\rm}\fi
\def\tenpoint{\def\rm{\fam0\tenrm}
\textfont0=\tenrm \scriptfont0=\sevenrm \scriptscriptfont0=\fiverm
\textfont1=\teni  \scriptfont1=\seveni  \scriptscriptfont1=\fivei
\textfont2=\tensy \scriptfont2=\sevensy \scriptscriptfont2=\fivesy
\textfont\itfam=\tenit \def\it{\fam\itfam\tenit}\def\footnotefont{\ninepoint}%
\textfont\bffam=\tenbf \def\bf{\fam\bffam\tenbf}\def\sl{\fam\slfam\tensl}\rm}
\font\ninerm=cmr9 \font\sixrm=cmr6 \font\ninei=cmmi9 \font\sixi=cmmi6
\font\ninesy=cmsy9 \font\sixsy=cmsy6 \font\ninebf=cmbx9
\font\nineit=cmti9 \font\ninesl=cmsl9 \skewchar\ninei='177
\skewchar\sixi='177 \skewchar\ninesy='60 \skewchar\sixsy='60
\def\ninepoint{\def\rm{\fam0\ninerm}
\textfont0=\ninerm \scriptfont0=\sixrm \scriptscriptfont0=\fiverm
\textfont1=\ninei \scriptfont1=\sixi \scriptscriptfont1=\fivei
\textfont2=\ninesy \scriptfont2=\sixsy \scriptscriptfont2=\fivesy
\textfont\itfam=\ninei \def\it{\fam\itfam\nineit}\def\sl{\fam\slfam\ninesl}%
\textfont\bffam=\ninebf \def\bf{\fam\bffam\ninebf}\rm}
%
%

\hyphenation{anom-aly anom-alies coun-ter-term coun-ter-terms}
\def\inv{^{\raise.15ex\hbox{${\scriptscriptstyle -}$}\kern-.05em 1}}

\def\Dsl{\,\raise.15ex\hbox{/}\mkern-13.5mu D} 
\def\dsl{\raise.15ex\hbox{/}\kern-.57em\partial}

 \def\Tr{{\rm Tr}}
\def\lspace{\ifx\answ\bigans{}\else\qquad\fi}
\def\lbspace{\ifx\answ\bigans{}\else\hskip-.2in\fi} 
\def\boxeqn#1{\vcenter{\vbox{\hrule\hbox{\vrule\kern3pt\vbox{\kern3pt
	\hbox{${\displaystyle #1}$}\kern3pt}\kern3pt\vrule}\hrule}}}
\def\mbox#1#2{\vcenter{\hrule \hbox{\vrule height#2in
		\kern#1in \vrule} \hrule}}  
%

\def\darr#1{\raise1.5ex\hbox{$\leftrightarrow$}\mkern-16.5mu #1}

\def\roughly#1{\raise.3ex\hbox{$#1$\kern-.75em\lower1ex\hbox{$\sim$}}}

\input epsf
\newif\ifdraft\draftfalse
\newif\ifinter\interfalse
\ifdraft\draftmode\else\interfalse\fi
\def\journal#1&#2(#3){\unskip, \sl #1\ \bf #2 \rm(19#3) }
\def\andjournal#1&#2(#3){\sl #1~\bf #2 \rm (19#3) }

\def\S{\hbox{$\bb S$}}

\def\frac#1#2{{#1\over#2}}

\def\inbar{\,\vrule height1.5ex width.4pt depth0pt}
\def\IC{\relax\hbox{$\inbar\kern-.3em{\rm C}$}}
\def\IR{\relax{\rm I\kern-.18em R}}
\def\IP{\relax{\rm I\kern-.18em P}}


%
\catcode`\@=11
\def\slash#1{\mathord{\mathpalette\c@ncel{#1}}}
\overfullrule=0pt

%
\def\draftnote#1{\ifdraft{\baselineskip2ex
                 \vbox{\kern1em\hrule\hbox{\vrule\kern1em\vbox{\kern1ex
                 \noindent \underbar{NOTE}: #1
             \vskip1ex}\kern1em\vrule}\hrule}}\fi}
\def\internote#1{\ifinter{\baselineskip2ex
                 \vbox{\kern1em\hrule\hbox{\vrule\kern1em\vbox{\kern1ex
                 \noindent \underbar{Internal Note}: #1
             \vskip1ex}\kern1em\vrule}\hrule}}\fi}


\newfam\frakfam
\font\teneufm=eufm10
\font\seveneufm=eufm7
\font\fiveeufm=eufm5
\textfont\frakfam=\teneufm
\scriptfont\frakfam=\seveneufm
\scriptscriptfont\frakfam=\fiveeufm

\def\bb{
\font\tenmsb=msbm10
\font\sevenmsb=msbm7
\font\fivemsb=msbm5
\textfont1=\tenmsb
\scriptfont1=\sevenmsb
\scriptscriptfont1=\fivemsb
}

\def\bbQ{\hbox{$\bb Q$}}

\def\bbZ{\hbox{$\bb Z$}}

\def\mmod#1{ [\![  #1 ] \!] }


\lref\AganagicSG{
  M.~Aganagic and S.~Shakirov,
  ``Knot Homology from Refined Chern-Simons Theory,''
  [arXiv:1105.5117 [hep-th]].
}

\lref\AldayAU{
  L.~F.~Alday, M.~Fluder and J.~Sparks,
  ``The Large N limit of M2-branes on Lens spaces,''
  JHEP {\bf 1210}, 057 (2012).
  [arXiv:1204.1280 [hep-th]].
}

\lref\AldayRS{
  L.~F.~Alday, M.~Bullimore and M.~Fluder,
  ``On S-duality of the Superconformal Index on Lens Spaces and 2d TQFT,''
  [arXiv:1301.7486 [hep-th]].
}

\lref\AldayKDA{
  L.~F.~Alday, M.~Bullimore, M.~Fluder and L.~Hollands,
  ``Surface defects, the superconformal index and q-deformed Yang-Mills,''
  [arXiv:1303.4460 [hep-th]].
}

\lref\BeniniMZ{
  F.~Benini, Y.~Tachikawa and B.~Wecht,
  ``Sicilian gauge theories and N=1 dualities,'' 
  JHEP {\bf 1001}, 088 (2010).
  [arXiv:0909.1327 [hep-th]].
}

\lref\BeniniNC{
  F.~Benini, T.~Nishioka and M.~Yamazaki,
  ``4d Index to 3d Index and 2d TQFT,''
  Phys.\ Rev.\ D {\bf 86}, 065015 (2012).
  [arXiv:1109.0283 [hep-th]].
}

\lref\BeniniNDA{
  F.~Benini, R.~Eager, K.~Hori and Y.~Tachikawa,
  ``Elliptic genera of two-dimensional N=2 gauge theories with rank-one gauge groups,''
  [arXiv:1305.0533 [hep-th]].
}

\lref\cherednik{
  I.~Cherednik,
  ``Double Affine Hecke Algebra,''
  Cambridge University Press, 2005
}

\lref\cheredniknon{
  I.~Cherednik, 
  ``Nonsymmetric Macdonald polynomials,''
  Internat.\ Math.\ Res.\ Notices {\bf 10}
   483 (1995).
}

\lref\cherednikMac{
  I.~Cherednik,
  ``Double affine {H}ecke algebras and {M}acdonald's conjectures,''
  Ann.\ of Math.\ {\bf 141} (1995).
}

\lref\CherednikNR{
  I.~Cherednik,
  ``Jones polynomials of torus knots via DAHA,''
  [arXiv:1111.6195 [math.QA]].
}

\lref\FrancoMM{
  S.~Franco,
  ``Bipartite Field Theories: from D-Brane Probes to Scattering Amplitudes,''
  [arXiv:1207.0807 [hep-th]].
}
   
\lref\GaddeIK{
  A.~Gadde, L.~Rastelli, S.~S.~Razamat and W.~Yan,
  ``The 4d Superconformal Index from q-deformed 2d Yang-Mills,''
  Phys.\ Rev.\ Lett.\  {\bf 106}, 241602 (2011).
  [arXiv:1104.3850 [hep-th]].
}

\lref\WittenNV{
  E.~Witten,
  ``Supersymmetric index in four-dimensional gauge theories,''
Adv.\ Theor.\ Math.\ Phys.\  {\bf 5}, 841 (2002).
[hep-th/0006010].
}

\lref\GaddeKB{
  A.~Gadde, E.~Pomoni, L.~Rastelli and S.~S.~Razamat,
  ``S-duality and 2d Topological QFT,''
  JHEP {\bf 1003}, 032 (2010).
  [arXiv:0910.2225 [hep-th]].
}

\lref\GaddeDDA{
  A.~Gadde and S.~Gukov,
  ``2d Index and Surface operators,''
  [arXiv:1305.0266 [hep-th]].
}

\lref\GaddeUV{
  A.~Gadde, L.~Rastelli, S.~S.~Razamat and W.~Yan,
  ``Gauge Theories and Macdonald Polynomials,''
Commun.\ Math.\ Phys.\  {\bf 319}, 147 (2013).
[arXiv:1110.3740 [hep-th]].
}
\lref\GangWY{
  D.~Gang,
  ``Chern-Simons theory on L(p,q) lens spaces and localization,''
  [arXiv:0912.4664 [hep-th]].
}

\lref\GaiottoWE{
  D.~Gaiotto,
  ``N=2 dualities,''
  JHEP {\bf 1208}, 034 (2012).
  [arXiv:0904.2715 [hep-th]].
}

\lref\GaiottoHG{
  D.~Gaiotto, G.~W.~Moore and A.~Neitzke,
  ``Wall-crossing, Hitchin Systems, and the WKB Approximation,''
  [arXiv:0907.3987 [hep-th]].
}

\lref\GaiottoXA{
  D.~Gaiotto, L.~Rastelli and S.~S.~Razamat,
  ``Bootstrapping the superconformal index with surface defects,''
  [arXiv:1207.3577 [hep-th]].
}

\lref\GorskyMK{
  E.~Gorsky, A.~Oblomkov, J.~Rasmussen and V.~Shende,
  ``Torus knots and the rational DAHA,''
  [arXiv:1207.4523 [math.RT]].
}

\lref\KallenNY{
  J.~Kallen,
  ``Cohomological localization of Chern-Simons theory,''
  JHEP {\bf 1108}, 008 (2011).
  [arXiv:1104.5353 [hep-th]].
}

\lref\OhtaEV{
  K.~Ohta and Y.~Yoshida,
  ``Non-Abelian Localization for Supersymmetric Yang-Mills-Chern-Simons Theories on Seifert Manifold,''
  Phys.\ Rev.\ D {\bf 86}, 105018 (2012).
  [arXiv:1205.0046 [hep-th]].
}

\lref\KapustinKZ{
  A.~Kapustin, B.~Willett and I.~Yaakov,
  ``Exact Results for Wilson Loops in Superconformal Chern-Simons Theories with Matter,''
  JHEP {\bf 1003}, 089 (2010).
  [arXiv:0909.4559 [hep-th]].
}

\lref\KinneyEJ{
  J.~Kinney, J.~M.~Maldacena, S.~Minwalla and S.~Raju,
  ``An Index for 4 dimensional super conformal theories,''
  Commun.\ Math.\ Phys.\  {\bf 275}, 209 (2007).
  [hep-th/0510251].
}

\lref\Macdonaldnon{
  I.~Macdonald,
  ``Affine {H}ecke algebras and orthogonal polynomials,''
  Ast\'erisque {\bf 237}, 189 (1996).
}

\lref\MacdonaldBook{
  I.~G.~Macdonald,
  {\it Symmetric functions and Hall polynomials},
  Oxford, 1979.
}

\lref\MacdonaldHecke{
  I.~G.~Macdonald,
  {\it Affine Hecke algebras and orthogonal polynomials},
  Cambridge, 2003.
}

\lref\Marshall{
  D.~Marshall,
  ``{Symmetric and nonsymmetric Macdonald polynomials},''
  Ann.\ Comb.\ {\bf 3} 385 (1999).
}

\lref\Noumi{
  M.~Noumi,
  ``Affine Hecke algebras and Macdonald polynomials,''
  Progr.\ Math.\  {\bf 160} 365 (1998).
}

\lref\RomelsbergerEC{
  C.~Romelsberger,
  ``Calculating the Superconformal Index and Seiberg Duality,''
  [arXiv:0707.3702 [hep-th]].
}

\lref\RomelsbergerEG{
  C.~Romelsberger,
  ``Counting chiral primaries in N = 1, d=4 superconformal field theories,''
  Nucl.\ Phys.\ B {\bf 747}, 329 (2006).
  [hep-th/0510060].
}

\lref\XieMR{
  D.~Xie and M.~Yamazaki,
  ``Network and Seiberg Duality,''
  JHEP {\bf 1209}, 036 (2012).
  [arXiv:1207.0811 [hep-th]].
}

\lref\YamazakiFVA{
  M.~Yamazaki,
  ``Four-dimensional superconformal index reloaded,''
  Theor.\ Math.\ Phys.\  {\bf 174}, 154 (2013), [Teor.\ Mat.\ Fiz.\  {\bf 174}, 177 (2013)].
}

\lref\KapustinKZ{
  A.~Kapustin, B.~Willett and I.~Yaakov,
  ``Exact Results for Wilson Loops in 
  Superconformal Chern-Simons Theories with Matter,''
  JHEP {\bf 1003}, 089 (2010).
  [arXiv:0909.4559 [hep-th]].
}

\lref\FestucciaWS{
  G.~Festuccia and N.~Seiberg,
  ``Rigid Supersymmetric Theories in Curved Superspace,''
  JHEP {\bf 1106}, 114 (2011).
  [arXiv:1105.0689 [hep-th]].
}

\lref\JafferisUN{
  D.~L.~Jafferis,
  ``The Exact Superconformal R-Symmetry Extremizes Z,''
  JHEP {\bf 1205}, 159 (2012).
  [arXiv:1012.3210 [hep-th]].
}

\lref\HamaAV{
  N.~Hama, K.~Hosomichi and S.~Lee,
  ``Notes on SUSY Gauge Theories on Three-Sphere,''
  JHEP {\bf 1103}, 127 (2011).
  [arXiv:1012.3512 [hep-th]].
}

\lref\KimWB{
  S.~Kim,
  ``The Complete superconformal index for N=6 Chern-Simons theory,''
  Nucl.\ Phys.\ B {\bf 821}, 241 (2009), 
  [Erratum-ibid.\ B {\bf 864}, 884 (2012)].
  [arXiv:0903.4172 [hep-th]].
}

\lref\ImamuraSU{
  Y.~Imamura and S.~Yokoyama,
  ``Index for three dimensional superconformal field theories with general R-charge assignments,''
  JHEP {\bf 1104}, 007 (2011).
  [arXiv:1101.0557 [hep-th]].
}

\lref\BeniniUI{
  F.~Benini and S.~Cremonesi,
  ``Partition functions of N=(2,2) gauge theories on $S^2$ and vortices,''
  [arXiv:1206.2356 [hep-th]].
}

\lref\DoroudXW{
  N.~Doroud, J.~Gomis, B.~Le Floch and S.~Lee,
  ``Exact Results in D=2 Supersymmetric Gauge Theories,''
  [arXiv:1206.2606 [hep-th]].
}

\lref\WRLens{
  S.~S.~Razamat and B.~Willett,
  ``Global Properties of Supersymmetric Theories and the Lens Space,''
[arXiv:1307.4381 [hep-th]].
}

\lref\SeibergPQ{
  N.~Seiberg,
  ``Electric - magnetic duality in supersymmetric nonAbelian gauge theories,''
  Nucl.\ Phys.\ B {\bf 435}, 129 (1995).
  [hep-th/9411149].
}

\lref\SpiridonovZA{
  V.~P.~Spiridonov and G.~S.~Vartanov,
  ``Elliptic Hypergeometry of Supersymmetric Dualities,''
  Commun.\ Math.\ Phys.\  {\bf 304}, 797 (2011).
  [arXiv:0910.5944 [hep-th]].
}

\lref\DolanQI{
  F.~A.~Dolan and H.~Osborn,
  ``Applications of the Superconformal Index for Protected Operators and q-Hypergeometric Identities to N=1 Dual Theories,''
  Nucl.\ Phys.\ B {\bf 818}, 137 (2009).
  [arXiv:0801.4947 [hep-th]].
}

\lref\Rains{
  E.~M.~Rains,
  ``{Transformations of elliptic hypergeometric integrals},''
  Ann.\ of Math.\ {\bf 171},
  169 (2010).
  [math.QA/0309252].
}

\lref\SpiridonovHF{
  V.~P.~Spiridonov and G.~S.~Vartanov,
  ``Elliptic hypergeometry of supersymmetric dualities II. Orthogonal groups, knots, and vortices,''
  [arXiv:1107.5788 [hep-th]].
}

\lref\Yamazaki{
  M.~Yamazaki, {\it in progress}.
}

\lref\MekareeyaTN{
  N.~Mekareeya, J.~Song and Y.~Tachikawa,
  ``2d TQFT structure of the superconformal indices with outer-automorphism twists,''
  JHEP {\bf 1303}, 171 (2013).
  [arXiv:1212.0545 [hep-th]].
}

\lref\LemosPH{
  M.~Lemos, W.~Peelaers and L.~Rastelli,
  ``The Superconformal Index of Class S Theories of Type D,''
  [arXiv:1212.1271 [hep-th]].
}

\lref\KirillovNoumi{
  A.~N.~Kirillov  and M.~Noumi,
  ``Affine Hecke algebras and raising operators for Macdonald
  polynomials,''
  Duke Math.\ J.\ {\bf 93} 1 (1998).
}


\rightline{PUPT-2445}
\rightline{IPMU13-0115}

\Title{
\vbox{\baselineskip12pt \hbox{}}}
{\vbox{\centerline{S-duality and the ${\cal N}=2$ Lens Space Index}}}

\centerline{Shlomo S.~Razamat$^1$ and Masahito Yamazaki$^{{2,3}}$}
\bigskip
  \centerline{$^{1}${\it Institute for Advanced Study, Princeton, NJ 08540, USA}}
\vskip .1in
 \centerline{$^{2}${\it Princeton Center for Theoretical Science,
 Princeton University, Princeton, NJ 08544, USA}}
\vskip .1in
 \centerline{$^{3}${\it Kavli Institute for the Physics and Mathematics of the Universe (WPI),}}
 \centerline{\it University of Tokyo, Kashiwa, Chiba 277-8583, Japan}

\vskip .4in
\centerline{\bf Abstract}

\noindent We discuss some of the analytic properties of lens space indices for
4d ${\cal N}=2$ theories of class ${\cal S}$.
The S-duality properties of these theories
highly constrain the lens space indices, and 
imply in particular that 
they are naturally acted upon by 
a set of commuting difference operators corresponding
to surface defects.
We explicitly identify the difference operators
to be a matrix-valued generalization of 
the elliptic Ruijsenaars-Schneider model.
In a special limit these difference operators can be 
expressed naturally in terms of Cherednik operators appearing in the 
double affine Hecke algebras, with the eigenfunctions  given by
non-symmetric Macdonald polynomials.

\vfill

\Date{June 2013}

\newsec{Introduction}

The supersymmetric index~\refs{\KinneyEJ,\RomelsbergerEG}, {\it a.k.a.}\
the twisted partition function on $\S^3\times \S^1$, 
is a powerful tool to extract quantitative data about
strongly coupled superconformal field theories (SCFTs).
While the lack of small parameters prevents us from performing
direct computations at the IR fixed point,
the indices can often be computed in the UV, where
we have a known Lagrangian description of the physics.
Since the index is independent of
gauge couplings~\refs{\RomelsbergerEC,\FestucciaWS},
we can then identify the supersymmetric indices computed from the UV description
 with the  superconformal indices at the IR fixed point. Following this logic
the superconformal indices have been computed for many different
theories in various dimensions,
and have provided impressive tests of
non-perturbative dualities. For example \DolanQI,  the equality of supersymmetric
indices for Seiberg-dual pairs 
in 4d   \SeibergPQ\
is due to remarkable identities of special functions  
appearing in \Rains\ (see also \SpiridonovZA, \SpiridonovHF).

In this paper we will be interested in 4d ${\cal N}=2$ SCFTs of class ${\cal S}$
\refs{\GaiottoWE, \GaiottoHG} which are obtained 
 by compactifying the 6d $(2,0)$ 
theory of type $A_{N-1}$ on punctured Riemann surfaces ${\cal C}$.
In most of these examples with $N>2$ there are no known Lagrangian
descriptions of these theories even in the UV.
However, the indices of these theories are severely constrained by their symmetries
and interrelations. The symmetry which is most useful here is S-duality:
the index should be invariant
 under marginal deformations and thus should be the same when computed 
in any one of the duality frames.\foot{It follows 
from this that the index defines a 3-parameter family of 
2d TQFTs on ${\cal C}$ \GaddeKB. This TQFT in a 1-parameter slice
coincides with the 2d $q$-deformed Yang-Mills 
theory in the zero-area limit \GaddeIK\ .}
In fact, it has been pointed out in \GaiottoXA\ that 
the assumption of 4d ${\cal N}=2$ S-duality is powerful enough to 
completely determine the superconformal indices, and 
leads to  manifestly S-duality invariant expressions for them
(obtained previously in \GaddeUV).

Here we will discuss  another interesting twist of this story.
We will be interested in studying the lens space index \BeniniNC, a 
twisted partition function on $\S^3/{\bbZ}_r\times \S^1$.
This is a generalization of the ordinary superconformal indices ($r=1$),
and has some new features not present in
their $r=1$ counterparts. The lens index of a gauge theory is determined as a 
sum over the integer holonomies
{\it i.e.}, discrete Wilson lines parametrized by $(\bbZ_r)^{N-1}$. 
Moreover, one can turn on non-trivial holonomies  for global symmetries.
The lens index is thus a function of fugacities for the global symmetries of the 
theory, {\it and} of the discrete holonomies for those symmetries.
With this extra structure the lens index contains more refined  information
about the IR fixed points than the ordinary superconformal
indices.\foot{
In particular the lens index is sensitive to the global structure of the group,
unlike the $r=1$ index, and can be used to distinguish dualities differing by such global properties~\refs{\WittenNV,\WRLens}
: this fact however will not be important to us. 
Moreover, 
some 
of the known exactly-localized partition functions, such as 
3d $S^3$ partition function (\KapustinKZ, \JafferisUN, \HamaAV), 
3d lens space partition function (\GangWY, \KallenNY, \OhtaEV,
\AldayAU), 
3d $S^1\times S^2$ index (\KimWB, \ImamuraSU),
and 2d $S^2$ partition function (\BeniniUI, \DoroudXW), are believed
to be deducible from a suitable reduction of the
4d lens indices \refs{\BeniniNC,\YamazakiFVA}.
}  

Our goal thus will be to outline some of the features of the lens
indices of theories of class ${\cal S}$ which can be deduced by exploiting 
their S-duality properties. Recently, the authors of \AldayRS\ have verified that,
 at least  in certain limits and for $A_1$ quivers,
the lens space indices of theories of class ${\cal S}$ are consistent with S-dualities
 of these theories. Our approach will be however orthogonal to this: we will {\it assume}
that the lens index of all theories of class ${\cal S}$ is independent of the 
S-duality frame it is computed in, and will discuss what  properties of the index follow from this assumption. 
In this way we thus will be able to say something about lens indices of theories which do not have
any known Lagrangian description.

In particular  following the technology developed in \GaiottoXA\
for the $r=1$ case we will study the analytical properties of the lens index
as a function of certain flavor fugacities. We will show that a class of poles 
of these indices
can be easily deduced. 
Moreover the residues of these poles are encoded implicitly  
in certain difference operators. It then will follow that the lens index 
has simple form when written as a sum of eigenfunctions of these operators.
As argued in \GaiottoXA, such residue computations
are related to  RG flows triggered by  turning on space-time dependent VEVs.
The residue of the index then describes the index of the IR theory in presence of such a VEV.
In general turning on space-time dependent VEVs will result in the IR theory having extended 
defects which in our setup are surface defects. One should thus
view the difference
operators obtained in the procedure of \GaiottoXA\ as  introducing certain surface defects
into the index computation.\foot{See~\GaddeDDA\ for a different, more direct, computation of the indices
of such surface defects.}

The difference operators we will obtain depend on three parameters ($p, q$, and $t$), the ${\cal N}=2$ superconformal
fugacities,
and act non-locally on the lattice $(\bbZ_r)^{N-1}$
parameterizing the integer holonomies for a global symmetry.
Mathematically these operators can be thought of as matrix-valued generalizations of 
elliptic Ruijsenaars-Schneider operators. S-duality is translated into a number of mathematical 
properties satisfied by the difference operators, such as the commutativity 
and self-adjointness under the vector multiplet measure.\foot{We would like to 
urge the more mathematically-oriented readers to prove these properties explicitly.}

In a certain limit of the superconformal fugacities, $p=0$,  similar to the  Macdonald limit for the $r=1$ 
case \refs{\GaddeUV,\GaiottoXA},
 we find that our difference operators can be related to a well-studied structure in mathematics. Namely,
these difference operators are related (by conjugation)
 to a symmetric combination of the Cherednik operators of
the double affine Hecke algebra (DAHA) \cherednik,
and their eigenfunctions are given by {\it non-symmetric} Macdonald polynomials studied for example in
\cherednik, \MacdonaldHecke.

\

This paper is organized as follows.
After a brief summary of the lens space indices for ${\cal N}=2$ theories 
in Sec.\ 2, we discuss the general strategy of computing poles and residues of the lens index
in Sec.\ 3.
We then analyze the difference operators, their eigenfunctions
and the lens space indices in more detail in Sec.\ 4.
We will also discuss two simplifying limits, namely the $p=0$ limit (Sec.\ 5)
and  $r\to \infty$ limit (Sec.\ 6).
Finally we make some further comments on our results in  Sec.\ 7.
Several appendices include technical details and developments.

\newsec{Lens Space Index}

Let us first briefly review the 4d ${\cal N}=2$  lens space index,
the supersymmetric partition function on $\S^3/{\bbZ}_r\times
\S^1$~\BeniniNC.
The lens space $L(r,1)=\S^3/{\bbZ}_r$ is given by the following
discrete identification
on $S^3$
\eqn\lens{
(z_1,\,z_2)\sim (e^{\frac{2\pi i}{r}}\,z_1,\, e^{-\frac{2\pi
i}{r}}\,z_2)\,, \qquad |z_1|^2+|z_2|^2=1\,.
}  This orbifold acts on the Hopf fiber 
of $\S^3$:  ${\bbZ}_r\subset U(1)_1\subset SU(2)_1 \subset SU(2)_1\times SU(2)_2\sim SO(4)$.
For the special case $r=1$ we recover the round sphere $\S^3$,
whereas in the opposite limit $r\to \infty$ the Hopf fiber shrinks and we obtain $L(r\to \infty, 1)\sim \S^2$.

The 4d ${\cal N}=2$ lens index is defined 
as\foot{The fugacities $p,q,t$ in our paper are related to 
$t_{\rm there},y_{\rm there}, v_{\rm there}$ of
 \BeniniNC\ as
$
p=t_{\rm there}^3\, y_{\rm there},\quad 
q=t_{\rm there}^3\, y_{\rm there}^{-1}, \quad
t=t^4_{\rm there}\, v_{\rm there}^{-1}.
$
We have denoted the $\bbZ_{p_{\rm there}}$ orbifold action in \BeniniNC\
by $\bbZ_r$, in order to save the notation $p$ for fugacity.
} 
\eqn\Idef{
{\cal I}(p,q,t;a)={\Tr} \left[ 
(-1)^F\,\left(\frac{t}{pq}\right)^r\,
 p^{j_{2}+j_1}\,
 q^{j_{2}-j_1}\,
 t^{R}\,
 \prod_i a_i^{f_i} 
 \right]
 \,,
} 
where the trace is over the Hilbert space on $\S^3/{\bbZ}_r$, $F$ the fermion number, $j_1, j_2$ the Cartans of the rotation group $SU(2)_1\times
SU(2)_2\sim SO(4)$, $R$ the $U(1)$ generator of $SU(2)_R$ R-symmetry and 
$r$ the generator of $U(1)_R$, and $f_i$ the flavor $U(1)$ symmetry (if there are any).
The index depends on the superconformal  fugacities $(p,q,t)$ and the fugacities for
flavor symmetries, the $a_i$'s.
We assume that the fugacities satisfy
the following conditions:
\eqn\fugineq{
|p|, |q|, |t| < 1\, , \quad |t|>|pq| \, , \quad |a_i|=1 \, .
}
This ensures the convergence of the definition \Idef, and 
will be important for the residue calculus in the next section.
In our residue calculus we will analytically continue the index by taking some
of the fugacities $a_i$ to be more general while keeping the rest on the unit circle.

The definition \Idef\ of the lens index is similar to the ordinary superconformal index ($r=1$).
However there is one qualitatively new feature which one should  consider for $r>1$:
we should (can) turn on non-trivial discrete Wilson lines (holonomies) $V$ for the gauge (flavor) 
vector fields,
since $\pi_1(\S^3/{\bbZ}_r)={\bbZ}_r$ and thus is non-trivial.
For a simply-connected gauge group this is
parameterized by elements in the Cartan of the gauge group $G$
\eqn\V{
V={\rm diag}(
e^{\frac{2\pi i m_1}{r}} , 
e^{\frac{2\pi i m_2}{r}} , \cdots , 
e^{\frac{2\pi i m_N}{r}} 
) \, ,
}
where the integers $m_i$'s take values in 
$\bbZ_r$. In this paper we will be interested in  $G=SU(N)$ and then we
also have $\sum_i m_i=0$
modulo $r$.
The holonomies satisfy $V^r=1$ since the $r$th power of this discrete Wilson line is contractible.
In presence of the holonomies  the gauge group is broken.
For  $G=SU(N)$  we have
\eqn\gaugebreaking{
SU(N)\to S\left[ \prod_{i=1}^N U(N_i) \right] \,,\qquad  \sum_{i=1}^N N_i=N \,,
}
where we defined 
\eqn\Nidef{
N_i:=\# \{  1\le j\le N \,|\, m_j=i \} \,,
} and we defined $U(0)$
to be the trivial group.
The Hilbert space factorizes into 
sectors with different values of $m$'s, and 
the lens index is defined as a sum over
different holonomy sectors specified by $m$.
We will see that this subtlety modifies the discussion of 
\GaiottoXA\ in an interesting way, and generalizes the 
mathematical structures behind the usual ($r=1$) superconformal index.

\

Given these general definitions one can compute the lens index 
of different multiplets.
The 4d ${\cal N}=2$ $A_{N-1}$ theories of class ${\cal S}$ 
are constructed from
two basic ingredients: the trinion theory and the vector multiplet associated
with cylinders \GaiottoWE. The trinion theory depends on the types of punctures,
 and is generically strongly-coupled. For the computations in  this paper we only 
need to know the lens index of bi-fundamental hypermultiplets, {\it i.e.}
the theory corresponding to a sphere with
two full punctures and one simple puncture.
This multiplet is in bifundamental representation of 
$SU(N)_{\mb b}\times SU(N)_{\mb z}$ flavor symmetries.
Moreover the half-hypers are charged (with opposite charges)  under a $U(1)_a$ symmetry.\foot{In 
the special case $N=2$ there is no distinction between full and
simple punctures. Since ${\bf 2}$ of $SU(2)$ is pseudoreal,
the hypermultiplet here can be decomposed into two half-hypermultiplets, and the
trinion theory is given by trifundamental half-hypermultiplets 
under global symmetries $SU(2)^3$.
}
The lens index of this theory, in addition to the superconformal fugacities $p,\,q,\,t$,
also depends on fugacities ${\mb b}$, ${\mb z}$, and $a$ for the global symmetry
$SU(N)_{\mb b}\times SU(N)_{\mb z}\times U(1)_a$.
Moreover, we can also turn on non-trivial holonomies for these symmetries. In
what follows we will need to consider only holonomies for $SU(N)_{\mb b}$ and $SU(N)_{\mb z}$
which we will denote by $\widetilde {\mb m}$ and ${\mb m}$ respectively.\foot{
In principle, one could also discuss adding a holonomy for the $U(1)$ symmetry under which the hyper-multiplet is charged. However, this adds complexity not needed for our discussion and thus we will 
refrain from doing so.
}
The lens index for this trinion theory is given by \refs{\BeniniNC}\
\eqn\trinionold{
\eqalign{
{\cal I}_{H}^{({\mb m},\, \widetilde {\mb m})}(a,{\mb b},{\mb z})=
{\cal I}^0_{H}\,\,
\prod_{s=\pm 1}
\prod_{i,j=1}^{N}
&\Gamma\left( t^{\frac{1}{2}} p^{\mmod{s(m_{i}+\widetilde m_{j})}} (z_i)^s (b_j)^s a^s ; pq,p^r
\right)  \cr
&\qquad
\times \Gamma\left( t^{\frac{1}{2}} q^{r-\mmod{s(m_{i}+\widetilde m_{j})}} (z_i)^s
 (b_j)^s a^s ; pq,q^r
\right) 
\, .
}
}
We have defined the elliptic gamma function
 $\Gamma(x;p,q)$ by
\eqn\ellGamma{
\Gamma(x;p,q)=\prod_{i,j\ge 0} \frac{1-x^{-1}\, p^{i+1} q^{j+1}}{1-x\, p^{i} q^{j}} \, .
}
For an integer $m$ we define $\mmod{m}$ to be $m$ modulo $r$,
{\it i.e.}, 
an integer $0\le \mmod{m}< r$ such that $m\equiv \mmod{m}$ modulo $r$.
The holonomies for $SU(N)$ flavor symmetries satisfy $\mmod{\sum_{i=1}^N \widetilde m_{i}}=\mmod{\sum_{i=1}^N m_{i}}=0$.

We will also need the lens index of the  ${\cal N}=2$ $SU(N)$ vector multiplet
\eqn\measureold{
\eqalign{
{\cal I}_V^{({\mb m})}({\mb z})={\cal I}_V^0\, \, 
&%
\left(
\frac{(p^r;p^r)}{
\Gamma(t ;pq, p^r) 
} \frac{(q^r;q^r)}{
\Gamma(t\, q^r ;pq, q^r) 
}
\right)^{N-1}
\prod_{1\le i<j\le N \, :\,  m_i=m_j} \left(1-\frac{z_i}{z_j} \right)^{-1} \left(1-\frac{z_j}{z_i} \right)^{-1}
\cr
&\times \prod_{i\ne j} 
\frac{1}{
\Gamma\left(t\, p^{\mmod{m_{i}\! -\! m_{ j}}} z_i/z_j ;pq, p^r\right) 
}
\frac{1}{
\Gamma\left(t\, q^{r-\mmod{m_{i}\! - \!m_{j}}} z_i/z_j;pq, q^r\right) 
}\cr
&\times \prod_{i\ne j} 
\frac{1}{
\Gamma\left( p^{\mmod{m_{i}\! -\! m_{ j}}} z_i/z_j ;pq, p^r\right) 
}
\frac{1}{
\Gamma\left( q^{r-\mmod{m_{i}\! - \!m_{j}}} z_i/z_j;pq, q^r\right) 
}
\, .
}
}
In the expressions \trinionold, \measureold, ${\cal I}_H^0$ and ${\cal I}_V^0$ are the zero-point contributions and are given by\foot{Note
that the zero-point contribution explicitly depends on the holonomies $m$, and 
is trivial in the case $r=1$, when we obtain the 
ordinary superconformal index. 
}
\eqn\zeropoint{
\eqalign{
{\cal I}_H^0
&=   \left(\frac{pq}{t}\right)^{\frac{1}{4} \left[\sum_{s=\pm1}\sum_{i,j=1}^N 
(\mmod{s(m_{i}+\widetilde m_{j})}-\frac{1}{r} \mmod{s(m_{i}+\widetilde m_{j})}^2)
\right]} \, , \cr
{\cal I}_V^0
&= 
 \left(\frac{pq}{t}\right)^{-\frac{1}{2} \left[\sum_{i,j=1}^N 
(\mmod{m^i-m^j}-\frac{1}{r} \mmod{m^i-m^j}^2) \right]}\, .
}
}
When we gauge a global symmetry,
we need to include  the index of the vector multiplet, ${\cal I}_V(z)$, sum over all the possible holonomies ${\mb m}$,  and integrate over the 
corresponding fugacity, with a measure given by
\eqn\measure{
[d{\mb z}]_{\mb m}=\frac{1}{\prod_{i=1}^N (N_i !)} \prod_{i=1}^{N-1} \frac{dz_i }{2\pi z_i} 
\prod_{1\le i<j\le N \, :\,  m_i=m_j} \left(1-\frac{z_i}{z_j} \right) \left(1-\frac{z_j}{z_i} \right) \ ,
}
which is the invariant Haar measure 
of the unbroken gauge group \gaugebreaking.
Note also that since $\prod_{i=1}^N z_i=1$, only $N-1$ of $z_i$'s are independent.
The integral over the $z_i$'s is performed over the contour $|z_i|=1$.


\newsec{Strategy}

A theory of class ${\cal S}$ corresponding to a Riemann surface ${\cal C}$ admits in general
several  descriptions. These descriptions correspond to different pair-of-pants decompositions
of the underlying Riemann surface. A given description is natural when certain couplings
are small ({\it i.e.} the corresponding tubes are long). Since the (lens) index is independent 
of the continuous couplings of the model, the  indices computed using different 
descriptions should agree.  This invariance of the index has far-reaching implications
for the form of the index.
In what follows we will deduce some of these implications. To do so 
we will follow the general strategy of \GaiottoXA.

\
 
Suppose we consider a theory corresponding to a Riemann surface ${\cal C}'$ which degenerates
into a trinion connected to the rest of the Riemann surface, ${\cal C}$, by a cylinder.
We also assume that two  of the punctures of the trinion
are full and one is simple (see Fig. 1).
\midinsert\bigskip{\vbox{{\epsfxsize=4.in
        \nobreak
    \centerline{\epsfbox{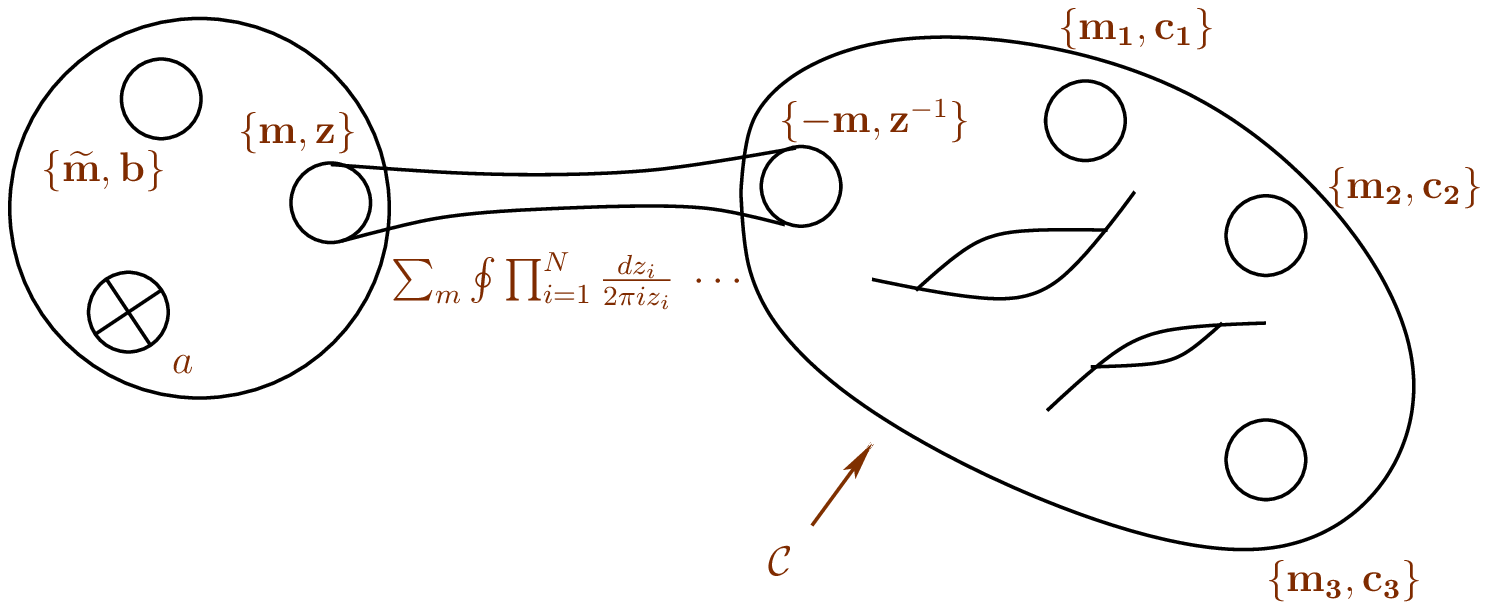}}
        \nobreak\bigskip
    {\raggedright\it \vbox{
\centerline{{\bf Fig 1.}
{\it  The trinion glued to a general Riemann surface ${\cal C}$
}}}}}}}
\bigskip\endinsert
Translated into the language of supersymmetric gauge theories, this
means that the theory ${\cal T}[{\cal C}']$ associated with the surface ${\cal C}'$
is obtained by gauging 
the diagonal $SU(N)$ symmetry inside $SU(N)^2$, 
one coming from the
trinion theory and another from the theory ${\cal T}[{\cal C}]$ for
the surface ${\cal C}$. The trinion is given by free $N^2$ ${\cal N}=2$ hypermultiplets
transforming under global symmetries $SU(N)_{\mb b}\times SU(N)_{\mb
z}\times U(1)_a$.
Let us turn on holonomies ${\mb m}$ for the  $SU(N)_z$ global
symmetry and 
$\widetilde {\mb m}$ for $SU(N)_b$.

The lens index of the theory ${\cal T}[{\cal C}']$ (denoted by ${\cal
I}$)
is obtained by
gluing that of the trinion (${\cal I}_H$) and 
of the theory ${\cal T}[{\cal C}]$ ($\widetilde {\cal I}$)
with a measure coming from the vector multiplet
\eqn\tmp{
\eqalign{
{\cal I}_{\mb \widetilde m}(a,{\mb b},-)= 
\sum_{{\mb m}}\oint  [d {\mb z}]_{{\mb m}} \,\, {\cal I}_{\rm H}^{({\mb m},\, \widetilde {\mb m})}(a, {\mb b}, {\mb z})\, \,{\cal I}_{V}^{({\mb m})}({\mb z})\,
\widetilde {\cal I}_{-\mb m}({\mb z}^{-1}, -) \, ,
}
}
where $-$ inside the arguments of  $\widetilde {\cal I}$
represents fugacities for the global symmetries associated with the remaining punctures of ${\cal C}$.
It should be emphasized here that the punctures of ${\cal C}$ other
than those associated with $SU(N)_{\mb z}$
are  arbitrary, and in particular theory ${\cal T}[{\cal C}]$ is in
general strongly-coupled with no known Lagrangian description;
 the only requirement for our computation 
is that ${\cal T}[{\cal C}']$ contains at least one minimal and one maximal puncture
so that we will be able to go to a description with a free bi-fundamental hypermultiplet 
coupled to ${\cal T}[{\cal C}]$.

\

In what follows we will study the poles $a=a^*$ of
the expression ${\cal I}_{\widetilde {\mb m}}(a, {\mb b},-)$
with respect to the fugacity $a$.
Physically a pole signifies that for a particular choice of fugacities 
a flat direction opens up, making the index divergent.
We can then trigger an RG flow by giving a VEV to the operator 
corresponding to the flat direction. This VEV in general will imply 
a non-trivial spatial profile for the operator and one can  argue
that  in IR the theory will be ${\cal T}[{{\cal
C}}]$ with a certain
surface defect. 
The residue is identified with the index 
of the  theory ${\cal T}[{{\cal
C}}]$ with such a surface defect \GaiottoXA.

We will find that  the residues of the poles of the index of ${\cal T}[{\cal C'}]$ 
in the $U(1)_a$ fugacity
are computed by certain
difference operators ${\cal O}_{a^*}$ acting on the index of ${\cal T}[{\cal C}]$:

\eqn\genresidue{
{\rm Res}_{a\to a^*} {\cal I}_{{\mb m}}(a,\,{\mb b},-)=
\sum_{{\mb n}\in (\bbZ_r)^{N-1}}
{{{\cal O}_{a^*}}^{\mb n}}_{{\mb m}}\; 
\widetilde {\cal I}_{{\mb n}}(\{\alpha_i\,b_i\},-)\,,
} where $\alpha_i$'s are monomials in fugacities $p$, $q$, and $t$. 
This difference operator 
in general acts non-locally 
on the discrete periodic $N-1$ dimensional lattice $(\bbZ_r)^{N-1}$
parametrizing the discrete Wilson lines
(which are denoted by ${\mb m, n}$ here).

We will discuss the difference operators 
in more detail in the next section, but let us here first explain the 
general implications of S-duality for the difference operators.

\

\subsec{\it The Implications of S-duality}

As we mentioned in the beginning of this section the fact that different 
descriptions of a theory of class ${\cal S}$ are interconnected by S-dualities  
implies that the lens index computed in different duality frames should be the same. 
In particular it should not matter which of the maximal punctures we decouple together 
with the minimal puncture corresponding to $U(1)_a$ in the residue computation above. 
Quantitatively this implies that when acting on the lens index of a theory of class ${\cal S}$
the difference operators satisfy
\eqn\sdualDiff{
\sum_{{\mb n}}
{{{\cal O}_{a^*}}^{\mb n}}_{{\mb m}}\; 
\widetilde {\cal I}_{{\mb n},\,{\mb m',\cdots}}(\{\alpha_i\,b_i\},\,{\mb b'},\dots)=
\sum_{{\mb n}}
{{{\cal O}_{a^*}}^{\mb n}}_{{\mb m'}}\; 
\widetilde {\cal I}_{{\mb m},\,{\mb n,\cdots}}({\mb b},\,\{\alpha_i\,b'_i\},\dots)\,.
}
Moreover the invariance of the index under S-duality implies~\GaiottoXA\ that the operators ${\cal O}_{a^*}$ should commute with each other for all the possible choices of $a^*$, and that
 ${\cal O}_{a^*}$ are self-adjoint 
with respect to the measure given by ${\cal I}_V^{(\mb m)}$ \measureold.
One can thus seek for a set of joint eigenfunctions, $\psi_\Lambda({\mb z};\,{\mb m})$,
 of all ${\cal O}_{a^*}$ which are
orthonormal under the natural vector multiplet measure appearing in the problem,
\eqn\measureN{
\sum_{{\mb n}}\,\oint[d{\mb z}]\; {\cal I}_V^{({\mb n})}({\mb z})\; 
\psi_{\Lambda}({\mb z};\,{\mb n})\;
\psi_{\widetilde \Lambda}({\mb z^{-1}};\,[-{\mb n}])
=\delta_{\widetilde \Lambda,\,\Lambda}\,.
} Given a set of such eigenfunctions one can, in principle, write the index of a theory of class ${\cal S}$ 
corresponding to a Riemann surface with genus $g$ and $s$ maximal punctures as
\eqn\genind{
{\cal I}=\sum_\Lambda \left(C_\Lambda\right)^{2g-2+s}\prod_{\ell=1}^s
\psi_{\Lambda}({\mb z}_\ell;\,{\mb n}_\ell)\,.
} In writing such an expression one assumes 
that the spectrum of the eigenvalues is non-degenerate.
This assumption is indeed correct in the $r=1$ case~\GaiottoXA.
However as we will see in next sections,
it is not true at least in certain limits of the lens index with $r>1$,
and the above ``diagonal''  form of the  index has to be modified 
to be ``block diagonal'' (see for example~\AldayRS).\foot{
One can expect that it should be possible to diagonalize also the ``blocks'':
the limits of  the parameters in which  the eigenfunctions are explicitly known,~\AldayRS\
and section 5 below, the lens index behaves in a somewhat subtle way so this statement was not explicitly checked.
} 
It is not unlikely that the structure constants
$C_\Lambda$ can be  also fixed by residue computations
 as was done in~\GaiottoXA\ for $r=1$ case.
Using S-duality one then in principle can translate the physical problem
 of finding the value of the lens index  to the mathematical
problem of finding the complete set of orthonormal eigenfunctions
for a set of commuting matrix-valued difference operators. 
To the best of our knowledge this mathematical problem however has
 not been solved yet for the difference operators at hand. 

\midinsert\bigskip{\vbox{{\epsfxsize=4.in
        \nobreak
    \centerline{\epsfbox{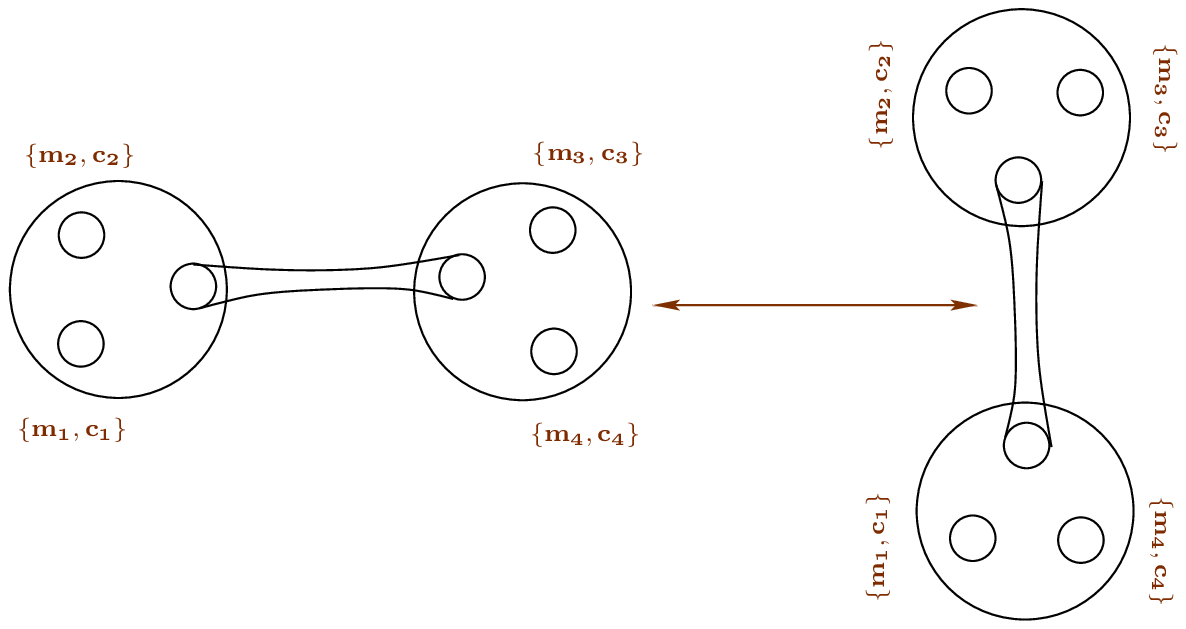}}
        \nobreak\bigskip
    {\raggedright\it \vbox{
\centerline{
{\bf Fig 1.}
{\it  The lens index is expected to be invariant under S-duality.
}
}
}}}}}
\bigskip\endinsert
 
\newsec{Residues and Difference Operators}

Let us  derive next the difference operator
\genresidue\ explicitly. We will make use of the formulas summarized in section 2.

\subsec{\it The Poles}

The loci of the poles of \tmp\ in the $U(1)_a$  fugacity can be deduced as follows.
 The index ${\cal I}_{\widetilde {\mb m}}(a,{\mb b},-)$ is computed by $z_i$ contour integrals. The integrands of these
integrals have numerous poles in $z_i$ with the position of the poles depending on various fugacities.
In particular when one varies these fugacities pairs of poles from opposite sides
of the integration contour can collide and pinch it: if all the contours are simultaneously pinched the
integrals giving ${\cal I}_{\widetilde {\mb m}}(a,{\mb b},-)$ diverge.\foot{If poles hit the integration
 contour without pinching  it no divergence occurs since in this case the contour can be smoothly
deformed away from the poles.}
Thus to find the loci of poles in $a$ one has to understand for which values of $a$ the contour integrals
in $z_i$ are simultaneously pinched.
Restricting to the case of $|a|<1$ we claim that this occurs when\foot{
The restriction $|a|<1$ corresponds to looking for poles coming from baryons.
There are also poles coming from anti-baryons which have $|a|>1$. Since these do not teach us anything new
we do not discuss them.  
}
\eqn\apolestwo{
a^*=t^{\frac12} (pq)^{\frac{n_1}{N}} p^{\frac{r n_2}{N}} q^{\frac{r n_3}{N}} \, ,
\quad n_1, n_2, n_3\ge 0 \, .
}
The poles for the $z_i$ integration which depend on the value of $a$ come
only  from the index ${\cal I}_{H}^{({\mb m},\, \widetilde {\mb m})}(a,{\mb b},{\mb z})$ of the trinion which we decoupled from
${\cal T}[{{\cal
C'}}]$ to obtain ${\cal T}[{{\cal
C}}]$. This index is given by \trinionold.
The poles of the integrand of \tmp\ coming from the index of the hypermultiplet
inside the integration contours are given by (recall \fugineq)
\eqn\polesIN{
\eqalign{
&z_i^{(q)}= \frac{1}{b_{\sigma(i)}\,a}\, t^{\frac12}\,q^{r\ell_i+\mmod{m_{i}+\widetilde m_{\sigma(i)}}}
(q\,p)^{s_i}\,,
\qquad \ell_i, s_i\geq0\,,\qquad i=1, \cdots,  N-1\,,\cr
&z_i^{(p)}=\frac{1}{b_{\sigma(i)}\,a}\, t^{\frac12}\,p^{r\ell_i+r-\mmod{m_{i}+\widetilde
m_{\sigma(i)}}}(q\,p)^{s_i}\,,\qquad \ell_i, s_i\geq0\,, \qquad
i=1, \cdots, N-1\, ,
}
} where we have introduced a permutation $\sigma\in S_N$. Note that $z_i^{(p)}$ and $z_i^{(q)}$ do not coincide for general values of $p$ and $q$.
 There are also two interesting sets of  poles outside the integration contours which come from terms with $z_N$ in the index of the decoupled trinion,
\eqn\polesOUT{
\eqalign{
&\tilde z_N^{(q)}=\frac{1}{\prod_{i=1}^{N-1}z_i}=\frac{1}{b_{\sigma(N)}\,a}\, t^{\frac12}\,q^{r\ell_N+\mmod{m_{N}+\widetilde m_{\sigma(N)}}}(qp)^{s_N}\,,\qquad \ell_N, s_N\geq0\,,\cr
&\tilde z_N^{(p)}=\frac{1}{\prod_{i=1}^{N-1}z_i}=\frac{1}{b_{\sigma(N)}\,a}\, t^{\frac12}\,p^{r\ell_N+r-\mmod{m_{N}+\widetilde m_{\sigma(N)}}}(qp)^{s_N}\,,\qquad \ell_N ,s_N\geq0\,.
}
} When a pole inside the contour coincides with one of
the poles outside, all the integration contours are
pinched at once 
and the whole integral has a pole.
Let us look for the poles of the form 
\eqn\apoles{
a=t^{\frac12}q^\alpha p^\beta\,.
} 
Then $\alpha$ and $\beta$ fit into one of the following four possibilities
\eqn\PQnew{
\eqalign{
&(q,q)\;:\;\qquad N \beta=\sum_{i=1}^Ns_i\,,\qquad
 N\alpha=N\beta+r\,\sum_{i=1}^{N}\ell_i+\sum_{i=1}^{N}\mmod{m_{i}+\widetilde m_{\sigma(i)}}\, , \cr
&(p,p)\;:\;\qquad N \alpha=\sum_{i=1}^Ns_i\,,\qquad N\beta=N\alpha+r\,\sum_{i=1}^{N}\ell_i+\sum_{i=1}^{N}\left(r-\mmod{m_{i}+\widetilde m_{\sigma(i)}}\right)\,  , \cr
&(q,p)\;:\;\qquad N\,\beta=\sum_{i=1}^Ns_i+r\,\ell_N+\left(r-\mmod{m_N+\widetilde m_{\sigma(N)}}\right)\,,\cr
&\qquad \qquad\qquad  N\,\alpha=\sum_{i=1}^Ns_i+r\sum_{i=1}^{N-1}\ell_i+\sum_{i=1}^{N-1}\mmod{m_i+\widetilde m_{\sigma(i)}}\,,\cr
&(p,q)\;:\;\qquad N\,\alpha= \sum_{i=1}^Ns_i+r\,\ell_N+\mmod{m_N+\widetilde m_{\sigma(N)}}\,,\cr
&\qquad \qquad\qquad  N\,\beta=\sum_{i=1}^Ns_i+r\sum_{i=1}^{N-1}\ell_i+\sum_{i=1}^{N-1}\left(r-\mmod{m_i+\widetilde m_{\sigma(i)}}\right)\,.
}
}
For example, a pole of the form $z_i^{(q)}$ and a pole $\tilde z_N^{(q)}$ coincide when
\eqn\look{
\eqalign{
\prod_{i=1}^{N-1}z_i^{(q)}=(\tilde z^{(q)}_N)^{-1}\, , 
}
}  
which leads to the case $(q,q)$ in \PQnew.
Similarly a pole of type $z_i^{(p)}$ coinciding with pole $\tilde
z_N^{(p)}$ results in case $(p,p)$ in \PQnew,
 and poles coming from $z_i^{(p)}$
($z_i^{(p)}$) coinciding  with  $\tilde z_N^{(q)}$ ($\tilde z_N^{(p)}$) result in
case $(q,p)$ ($(p,q)$)
in \PQnew. Thus from \PQnew\ and \apoles\ we derive \apolestwo.

\bigskip

The loci of the poles \apoles\  have a simple physical explanation in terms of surface defects.
For $r=1$ case discussed in \GaiottoXA, we have  poles  located at 
\eqn\apolesold{
a^*=t^{\frac{1}{2}} p^{\frac{n_1}{N}} q^{\frac{n_2}{N}} , \qquad n_1, n_2 \ge 0 \ .
}
The most basic pole is at $a^*=t^{\frac{1}{2}}$,
which corresponds to a VEV for a baryonic operator $B\sim Q^N$ built from the 
decoupled hypermultiplet. The two towers of poles in \apolesold\
 correspond then to VEVs for derivative operators $\partial_{12}^{n_1}
\partial_{34}^{n_2} B$. It was argued in \GaiottoXA\ that  such VEVs 
result in the IR theory having a certain surface defect.
 When we take ${\bbZ}_r$ orbifold \lens , the surviving states which are not charged
under global symmetries for which holonomies are turned on satisfy
$2j_1=0$ modulo $r$. 
This means that the surface defects are allowed only when $n_1-n_2=0$ modulo $r$.
After keeping only such poles from \apolesold, we find \apolestwo.

\

\subsec{\it The Residues}

Next we give an example of how to compute the residues at the poles~\apolestwo.
There are three ``basic'' poles: $a^*=t^{\frac12}(p\,q)^{\frac1N}, a^*=t^{\frac12}q^{\frac{r}{N}}$
and $a^*=t^{\frac12}q^{\frac{r}{N}}$. All the other poles are located at positions which are 
given by some product of these three.
Let us first quote the results for the residue at the  $a^*=t^{\frac12}(p\,q)^{\frac1N}$ 
\eqn\resbas{
\eqalign{
{\rm Res}_{a\to t^{\frac12}\,(p\,q)^{\frac{1}{N}}}& {\cal I}_{\mb m}(a,\,
{\mb b},\dots)=
\sum_{\mb n} \left[ {\cal O}_{a^*=t^{\frac12}(p\,q)^{\frac1N}}\right]^{\mb n}_{\,\, \mb m} \widetilde {\cal I}_{\mb n}(a,\,
{\mb b},\dots)
\cr
&=\sum_{I=1}^NF_I^{\{m\}}({\mb b})\,\,\widetilde {\cal I}_{\mb m}
\left(b_I\to b_{I}
(q\,p)^{\frac{1-N}{N}} , 
b_{i\ne I} \to (q\,p)^{\frac{1}{N} }  b_i \right) 
\cr &+ \sum_{I\neq J}^N
G^{\{m\}}_{I,J}({\mb b})\,\,\widetilde {\cal I}_{\{m_1,m_2,\dots ,m_I+1,\dots,m_J-1,\dots\}}
\Big(
b_I\to q^{\frac{1-N}{N}}\,p^{\frac1N}b_I
, \cr  
 & \qquad \qquad \qquad \qquad \qquad 
 b_J\to p^{\frac{1-N}{N}}\,q^{\frac1N}b_J,  \,
b_{i\ne I,J} \to (q\,p)^{\frac1N}b_i
\Big)\, ,
}
} 
One  can compute explcitly the functions $F$ and $G$.
For example, in the $A_1$ case with $a^*=t^{\frac{1}{2}} (p\, q)^{\frac{1}{N}}$ we obtain 
(no restrictions on $m$)
\eqn\FpoleTwo{
\eqalign{
&F^{m}_1=\left({\cal I}_V^{(m=0)} \right)^{-1}\,\frac{\Gamma((pq)^{\pm1};p^r,q^r)}{\Gamma(t^{\pm1};p^r,q^r)}\frac{\theta(q^{2m}\frac{t}{pq}b^{-2};q^r)\theta(p^{2m}\frac{pq}{t}b^{2};p^r)}{\theta(q^{2m}b^{-2};q^r)\theta(p^{2m}b^{2};p^r)}\,,\cr
&F^{m}_2=\left({\cal I}_V^{(m=0)} \right)^{-1}\,\frac{\Gamma((pq)^{\pm1};p^r,q^r)}{\Gamma(t^{\pm1};p^r,q^r)}\frac{\theta(q^{2m}\frac{pq}{t}b^{-2};q^r)\theta(p^{2m}\frac{t}{pq}b^{2};p^r)}{\theta(q^{2m}b^{-2};q^r)\theta(p^{2m}b^{2};p^r)}\,,\cr
&G^{m}_{1\,2}=
\left(\frac{p\,q}{t}\right)^{\frac{2+4m-r}{r}}\,
\left({\cal I}_V^{(m=0)} \right)^{-1}\,\frac{\Gamma((pq)^{\pm1};p^r,q^r)}{\Gamma(t^{\pm1};p^r,q^r)}
\frac{\theta(q^{2m}\frac{pq}{t}b^{-2};q^r)\theta(p^{2m}\frac{pq}{t}b^{2};p^r)}{\theta(q^{2m}b^{-2};q^r)\theta(p^{2m}b^{2};p^r)}\,,\cr
&G_{2\,1}^{m}=
\left(\frac{p\,q}{t}\right)^{\frac{2-4m+r}{r}}\,
\left({\cal I}_V^{(m=0)} \right)^{-1}\,\frac{\Gamma((pq)^{\pm1};p^r,q^r)}{\Gamma(t^{\pm1};p^r,q^r)}
\frac{\theta(q^{2m}\frac{t}{pq}b^{-2};q^r)\theta(p^{2m}\frac{t}{pq}b^{2};p^r)}{\theta(q^{2m}b^{-2};q^r)\theta(p^{2m}b^{2};p^r)}\,,
}
} 
where we denoted $(b_1,\,b_2)=(b,\,b^{-1}),\; (m_1, m_2)=(m, -m)$ and  defined 
\eqn\thetadef{
\theta(x;q):=(x;q) \left(\frac{q}{x};q\right)\, ,  \qquad (x;q):=\prod_{i=1}^{\infty} (1-x q^i)\ .
}
We also used the short-hand notation that $\pm$ in the argument of an expression represents the product of two instances of the expression with 
the plus sign and the minus sign in argument. For example
$\Gamma((pq)^{\pm 1};p^r, q^r)=\Gamma((pq)^{+1};p^r, q^r) \Gamma((pq)^{- 1};p^r, q^r)$.
Note that  \FpoleTwo\ are explicitly periodic in $m\sim
m+r$. Similar expressions can be obtained for the higher rank cases.

\

Let us now  derive \resbas.  Note first that  
the $z$-poles for $a^*=t^{\frac12}(p\,q)^{\frac1N}$ appear in the $(q,q)$ and $(q,p)$ sectors in \PQnew.
In $(q,q)$ sector we have to set $\ell_i=0$, $\widetilde m_{\sigma(i)}=r-m_i$ and  $s_I=1$ with $s_{i\neq I}=0$. The poles in $z_i$ which pinch the integration contours are located at
\eqn\zsoneone{
z_i=(q\,p)^{\delta_{iI}-\frac{1}{N}}\frac{1}{b_{\sigma(i)}}\,.
}
In $(q,p)$ sector we have to set $\ell_i=0$, $s_i=0$  for all $i$; for $i\neq I, N$ $\widetilde m_{\sigma(i)}=r-m_i$, and  $\widetilde m_{\sigma(I)}=r-m_I+1$, $\widetilde m_{\sigma(N)}=r-m_N-1$. The relevant poles 
in $z_i$ are located then at
\eqn\zsoneoneM{
z_{i\neq I,\,N}=(q\,p)^{-\frac{1}{N}}\frac{1}{b_{\sigma(i)}}\,,\qquad
z_I=q^{\frac{N-1}N}\,p^{-\frac{1}{N}}\,\frac{1}{b_{\sigma(I)}}\,,\qquad
z_N=p^{\frac{N-1}{N}}\,q^{-\frac{1}{N}}\frac{1}{b_{\sigma(N)}}\,.
} 
These two contributes gives the $F$ and $G$ terms in \resbas, respectively.

The difference operator computing the residue at $a^*= t^{\frac12}\,(p\,q)^{\frac{1}{N}}$
of the lens index with holonomy ${\mb m}$ involves 
 ``nearest neighbor'' points on the ${\mb m}$ lattice (Fig.~1). 
This is to be contrasted with the difference operator for a generic residue which 
will involve all points on the ${\mb m}$ lattice.

\

The $b$-dependent part of \FpoleTwo\ can be interpreted as counting the 2d
degrees of freedom localized on a surface defect.
 Let us take $F_1^m$ as an example. 
For $r=1$  $F_1^m$ involves factors of the form
\eqn\twoindex{
\frac{\theta(\frac{t}{pq}b^{-2};q)}{\theta(b^{-2};q)}\,, \qquad
\frac{\theta(\frac{pq}{t}b^{2};p)}{\theta(b^{2};p)}\,,
}
which have the form of elliptic genera of 2d multiplets \refs{\GaiottoXA,\GaddeDDA, \BeniniNDA}.
These indices are products and ratios of terms of the form $(1-b^{\pm 2} t^i p^j
q^k)$. Such terms survive the orbifold projection only when 
$j-k \pm 2m=0$ modulo $r$. 
This projection condition follows from the definition of the lens index \Idef\
and the ${\bbZ}_r$ action \lens; when translated along the Hopf fiber of $S^3/\bbZ_r$
the wavefunction  acquires a phase $(e^{\frac{2\pi
i}{N}})^{j-k}(e^{\frac{2\pi i}{N}})^{\pm 2m}$,
where the first factor comes from the spin $j_1$ 
and the second from the gauge field flux along the 
Hopf fiber (Aharanov-Bohm effect).
Keeping only such terms from the product in \twoindex, 
we obtain the combinations which appear in $F_1^m$
\eqn\twoindextwo{
\frac{\theta(q^{2m}\frac{t}{pq}b^{-2};q^r)}{\theta(q^{2m} b^{-2};q^r)}\,,\qquad
\frac{\theta(p^{2m}\frac{pq}{t}b^{2};q^r)}{\theta(p^{2m}b^{2};p^r)}\,.
}
\midinsert\bigskip{\vbox{{\epsfxsize=2.in
        \nobreak
    \centerline{\epsfbox{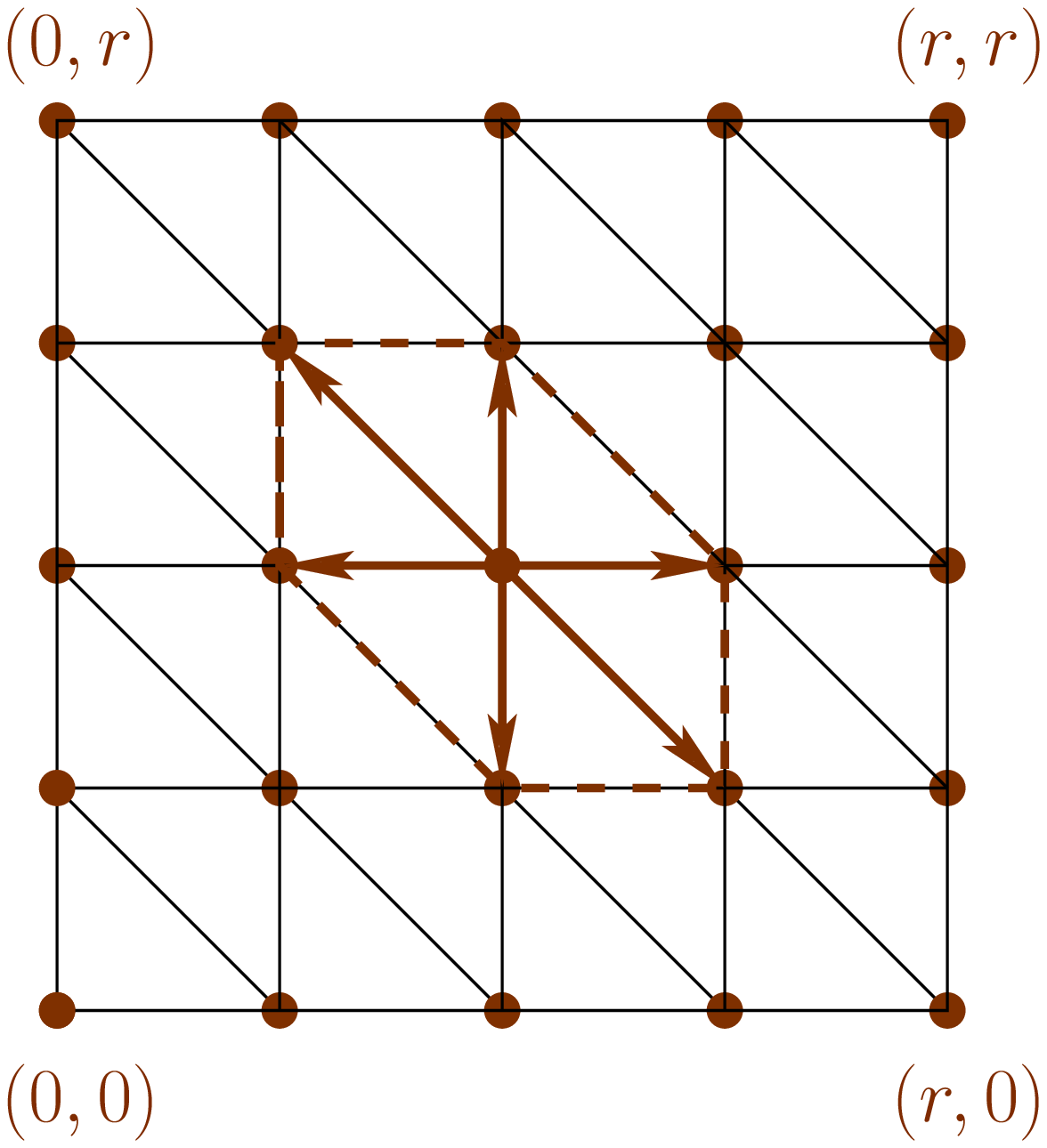}}
        \nobreak\bigskip
    {\raggedright\it \vbox{
{\bf Fig 1.}
{\it  The $(m_1,m_2)$ lattice of the $A_2$ case for $r=4$
(opposite sides are identified).
The red arrows represent the nearest neighbor sites for the difference operator 
computing residues at $a^*=t^{\frac12}(p\,q)^{\frac{1}{N}}$.
}}}}}}
\bigskip\endinsert

\

\noindent In the next sections we will discuss in more detail two simplifying limits
of the difference operators. 

\newsec{Macdonald Limit}

Let us discuss the limit of the index when one of the fugacities $p$ or $q$
 is vanishing. In what follows for concreteness we will take $p\to0$. For the $r=1$ case this limit is 
called the Macdonald index  \GaddeUV\ and we will keep this name also here:
this will be justified by the appearance of the
non-symmetric Macdonald polynomials.
A further limit of $p=0,\;t=q^r$ (Schur limit) for $A_1$ quivers was discussed in~\AldayRS.
Unlike the $r=1$ case where $p$
always appears in the index with non-negative powers,
the lens index for $r>1$ might contains
also negative powers of $p$: these powers come from the zero point energies \zeropoint.\foot{
In the $r=1$ case the power of $p$  couples to $\delta=\{{\cal Q}_{1+},\, {\cal Q}_{1+}^\dagger\}$ and thus
is  non-negative. However when $r>1$ the supercharge ${\cal Q}_{1+}$ does not correspond to a symmetry of the theory anymore due to the orbifold projection.
}
However, one 
can show (see~\AldayRS\ and/or appendix A here)
that the following quantity
\eqn\rescaledtrinion{
{\hat {\cal I}_H}(a,{\mb b},{\mb z}) \equiv \left({\cal I}_V^0({\bf m})\right)^{\frac12}\;
\left({\cal I}_V^0({\bf\widetilde m})\right)^{\frac12}\;{\cal I}_H(a,{\mb b},{\mb z}) \,,
}  has a well-defined limit as $p$ is taken to vanish. 
Moreover $\hat {\cal I}_{H}$ vanishes
in the limit unless there is a permutation $\hat \sigma\in {S}_N$ such that 
\eqn\condmac{
\forall\; i\qquad \mmod{m_i+\widetilde m_{\hat \sigma(i)}}=0\,.
}
In the rest of the section we will have in mind such a rescaled index,
and will drop the hat from the notations. Note that after the rescaling 
\rescaledtrinion\ one does not have to include the  zero-point energy in the vector multiplets.

\bigskip
The poles which survive in the limit 
are located at
\eqn\macpoles{
a=t^{\frac12}\,q^{\frac{r}{N}\,n}\,,\qquad n\ge 0 \,.
} 
The residue is computed to be
of the following form (for simplicity we assume here that all $m_i$ are different)
\eqn\macresidue{\eqalign{
&{\rm Res}_{a\to t^{\frac12}\,q^{\frac{r}N}}{\cal I}_{{\mb m}}(a,\,{\mb z})=
\sum_{\mb n} \left[ {\cal O}_{a^*=t^{\frac12}\,q^{\frac{r}N} }\right]^{\mb n}_{\,\, {\mb m}} \widetilde {\cal I}_{{\mb n}}(a,\,{\mb z}) \cr
&=\sum_{I=1}^N F_I({\mb z})\,\; \widetilde {\cal I}_{\mb m}(z_I\to q^{\frac{r(1-N)}{N}}\,z_{\hat \sigma(I)},\,
z_{i\neq I}\to q^{\frac{r}{N}}\,z_{\hat \sigma(i)})\cr
&+\sum_{I < J}G_{(I\,J)}({\mb z})
\,\; \widetilde {\cal I}_{\mb m}(
z_{I}\to q^{\frac{r}{N}-m_I+m_J}\,z_{\hat \sigma(J)},\,
z_{J}\to q^{\frac{r(1-N)}{N}-m_J+m_I}\,z_{\hat \sigma(I)},\, 
z_{i\neq I,\,J}\to q^{\frac{r}{N}}\,z_{\hat \sigma(i)})\cr
&+\sum_{I<J<K}G_{(I\,J\,K)}({\mb z})\, \;
 \widetilde {\cal I}_{\mb m}(
z_{I}\to q^{\frac{r}{N}-m_I+m_J}\,z_{\hat \sigma(J)},\,
z_{J}\to q^{\frac{r}{N}-m_J+m_K}\,z_{\hat \sigma(K)},\, \cr 
&\hskip 5cm z_{K}\to q^{\frac{r(1-N)}{N}-m_K+m_I}\,z_{\hat \sigma(I)},\,
z_{i\neq I,\,J}\to q^{\frac{r}{N}}\,z_{\hat \sigma(i)})\cr
&+
\cdots\,.
}
} It is straightforward to evaluate the functions $F_I$ and $G_{(I\, J\,\cdots)}$: we will quote the answer
for $A_1$ case momentarily (and for $A_2$ in appendix B). Note that in the Macdonald limit the
difference operators are local
on the lattice defined by ${\mb m}$ ({\it i.e.} the residue computed by this difference operator 
for the index with holonomy ${\mb m}$ act only on the index with the same value of
${\mb m}$). 

To derive \macresidue,  note that 
the pole \macpoles\
come only from the $(q,q)$ sector in \PQnew:
\eqn\macres{
a^*=t^{\frac12}\,q^{\alpha},\qquad\qquad \alpha=\frac{r}{N}\,\sum_{i=1}^{N}\ell_i+\frac{1}{N}\sum_{i=1}^{N}\mmod{m_{i}+\widetilde
m_{\sigma(i)}}\, ,
}
where $\sum_{i=1}^{N}\mmod{m_{i}+\widetilde m_{\sigma(i)}}$ is always divisible by $r$.
The permutation $\sigma$ introduced in Sec.\ 3 is in general different from
$\hat\sigma$ introduced above.
We will assume the generic scenario in which  all $m_i$ are different and  we 
order them such that $m_i>m_j$ if $i>j$. We will comment shortly on the case when this assumption 
does not hold.
 From~\condmac\ we deduce  that
$\widetilde m_{\hat \sigma(i)}=r-m_i$.
Let us evaluate the residues and the associated difference operators for the simplest case of $n=1$ in~\macpoles.
Here $\alpha=\frac{r}{N}$ which can be achieved either by setting  $\{\ell_I=1,\,\ell_{i\neq I}=0\}$ and $\sigma=\hat \sigma$, or by setting $\ell_i=0$ and 
choosing $\sigma\in {S}_N$ such that only for one value of $i$ $m_{\hat\sigma(i)}<m_{\sigma(i)}$.
 This can happen if $\sigma$ and $\hat \sigma$ differ by a single 
cycle of the form $(I_1\,I_2\,\cdots\, I_k)$ with $I_1<I_2<\cdots<I_k$.
The positions of the $z_i$ poles are thus given by
\eqn\pospolesmac{\eqalign{
&\{z_{I}=q^{\frac{r(N-1)}{N}}\,b_{\hat \sigma(I)}^{-1}\,,\quad
z_{i\neq I}=q^{-\frac{r}{N}}\,b_{\hat \sigma(i)}^{-1}\}
\,,\cr
&\{z_{I}=q^{-\frac{r}{N}+m_I-m_J}\,b_{\hat \sigma(J)}^{-1}\,,\quad
z_{J}=q^{\frac{r(N-1)}{N}+m_J-m_I}\,b_{\hat \sigma(I)}^{-1}\,,\quad
z_{i\neq I,\,J}=q^{-\frac{r}{N}}\,b_{\hat \sigma(i)}^{-1}\} \,,\cr
&\{z_{I}=q^{-\frac{r}{N}+m_I-m_J}\,b_{\hat \sigma(J)}^{-1}\,,\, \quad
z_{J}=q^{-\frac{r}{N}+m_J-m_K}\,b_{\hat \sigma(K)}^{-1}\,,\, \cr
& \hskip 4 cm  z_{K}=q^{\frac{r(N-1)}{N}+m_K-m_I}\,b_{\hat \sigma(I)}^{-1}\,,\,
z_{i\neq I,J,K}=q^{-\frac{r}{N}}\,b_{\hat \sigma(i)}^{-1}\}\,, \cr
&\cdots\,.
}
} Here  we assumed without loss of generality that
$I<J<K<\cdots$.  Collecting these contributions, we obtain \macresidue.
Finally, when  not all $m_i$ are different the terms $G_{(I_1\,\cdots I_k)}$ corresponding to cycles 
permuting equal masses are absent from the difference operator.

\subsec{$A_1$}

Let us  consider the $A_1$ quivers in more detail.
The $A_1$ case is special since 
$U(1)_a$ symmetry enhances to $SU(2)_a$, 
and the trinion theory is a tri-fundamental 
half-hypermultiplet under $SU(2)^3$, {\it i.e.}
there is no distinction here between minimal and maximal punctures.
To take advantage of this symmetry let us use the notation $\vec{z}=(z^1, z^2, z^3)=(a, b, z)$ 
and $\vec{m}=(m^1, m^2, m^3)=(m_a, m_b, m_c)$ where we also turned on a non-zero holonomy
for the $SU(2)_a$ symmetry.\foot{
The notation $\vec{m}=(m^1, m^2, \cdots )$ should not be confused with the previous notation
${\mb m}=(m_1, m_2, \cdots )$. The index for the former represents the three punctures of the trinion,
whereas the index for the latter represents the $N$ indices for the Cartan of the gauge group. 
In the the $A_1$ case here we have ${\mb m}_1=(m_1, -m_1)$ and ${\mb z}=(z,z^{-1})$.
}	
In this notation, the  Macdonald index of the trinion is given by
\eqn\Aonetrinion{
{\cal I}_H(z^1, z^2, z^3)= \prod_{\vec{s}=\pm 1} \frac{1}{\left(t^{\frac{1}{2}}\,q^{\mmod{-\vec{m} \cdot \vec{s}}}\,(z^1){}^{s^1}(z^2){}^{s^2}(z^3){}^{s^3};\,q^r\right)} \,.
}
The residue at $t^{\frac12}q^{\frac12}$ \macresidue\
in the  $A_1$ case  evaluates explicitly to
\eqn\nextres{
\eqalign{
&{\rm Res}_{a\to t^{\frac{1}{2}}q^{\frac{1}{2}r}}{\cal I}_{m}(a,b)=
\frac{1-t}{2\,(1-q^{-r})\,(q^r;q^r)(t;q^r)} \; {\cal H}_b\,\cdot \,\widetilde{\cal I}_m(b)
=\frac{(1-t)^2}{2\,(q^r;q^r)(t;q^r)}\times \cr
&\qquad\left( 
\frac{1}{(1-t)(1-q^{-r})}
\left[
\frac{1-t\,q^{2m-r}\,b^{-2}}{1-q^{r-2m}\,b^2}\,\widetilde{\cal I}_{m}(q^{\frac{1}{2}r}\,b)+
\frac{1-t\,q^{-2m}\,b^{2}}{1-q^{2m}\,b^{-2}}\,\widetilde{\cal I}_{m}(q^{-\frac{1}{2}r}\,b)\right]+\right.\cr
&\qquad\qquad\left.+\frac1{(1-q^{2m-r}\,b^{-2})(1-q^{-2m}\,b^{2})}\,\widetilde{\cal I}_{m}(q^{2m-\frac{1}{2}r}\,b^{-1})
\right)\,.
}
} 
The two terms on the second line come from $F_{1,2}$ and the term on the third line is $G_{(1\,2)}$. 
We have defined ${\cal H}_b$ as the difference operator computing the residue in this case.
A priori from the derivation of the previous sub-section 
this operator computes the residue when
 $m\neq 0,\frac{r}{2}$ since then the two $m_i$ are different.
However it is easy to show that the operator one obtains when
$m=0, \frac{r}{2}$  is equivalent to ${\cal H}_b$ when the latter acts on
symmetric functions: {\it i.e.} functions which are symmetric under the action of the Weyl group
which here is $f(b)=f(b^{-1})$. Indeed, when $m=0$ or $\frac{r}{2}$ the flavor group enhances 
from $S(U(1)\times U(1))$ to $SU(2)$ and the index is invariant under the Weyl group of $SU(2)$.

As we discussed in section 3 a consequence of the invariance of the lens index under S-duality is that 
\eqn\propSduality{
{\cal H}_{z^1}\, \widetilde{\cal I}_{m^1,m^2,..}(z^1,z^2,\dots)=
{\cal H}_{z^2}\,\widetilde {\cal I}_{m^1,m^2,..}(z^1,z^2,\dots)\,.
} Here  $\widetilde {\cal I}_{m^1,m^2,..}(z^1,z^2,\dots)$ is the lens index of a general $A_1$ theory
 of class ${\cal S}$.
 We can check explicitly for the  trinion \Aonetrinion\ that this property holds,
\eqn\invtrin{
\eqalign{
&{\cal H}_{z^1} \,\prod_{s_i=\pm1}
\frac{1}{\left(t^{\frac{1}{2}}\,q^{\mmod{-\vec{m}\cdot \vec{s}}}\,(z^1)^{s^1}(z^2)^{s^2}(z^3)^{s^3};\,q^r\right)}\propto\cr
&\qquad\qquad\prod_{s_i=\pm1}
\frac{1}{\left(\left(\frac{t}{q^r}\right)^{1/2}\,q^{\mmod{-\vec{m}\cdot \vec{s}}}\,(z^1)^{s^1}(z^2)^{s^2}(z^3)^{s^3};\,q^r\right)}
\times\cr
&\qquad\left[1-3\,t-3q^{-r}\,t^2+q^{-r}\,t^3-t\,(1+t)\,\sum_{i=1}^3((z^i)^2q^{-2m^i}+q^{-r}(z^i)^{-2}q^{2m^i}) \right.\cr
& \qquad \qquad+2\,t^{\frac32}\,q^{-\frac{r}2}\sum_{s^i=\pm1}
\prod_{i=1}^3(z^iq^{-m^i})^{s^i}+
2\, t^{\frac32}(q^{\frac{r}2}-q^{-\frac{r}2})z^1 z^2 z^3q^{-m^1-m^2-m^3}+
\cr 
&\qquad\qquad\left.+
2\, t^{\frac32}(q^{-\frac{3r}2}-q^{-\frac{r}2})(z^1 z^2 z^3)^{-1}q^{m^1+m^2+m^3}
\right]\,.
}
} Here we assumed for simplicity 
that the triplet $(m^1,m^2,m^3)$ satisfies strict triangle inequality, all $m_i$ are different and
satisfy $0<m_i\ll r$. The right-hand-side is explicitly symmetric in the three punctures although the operator acted only
on the first one. This fact can be viewed either as a non-trivial check of S-duality of $A_1$ quivers,
or if one takes S-duality for granted as a check of our technical procedure.
Note that the holonomies $m^i$ can be  absorbed into $z^i$ by redefining $\hat z^i= z^i\, q^{-m_i}$.
The only information about holonomies affecting the index is whether or not they satisfy certain exclusions. The three holonomies have to satisfy 
\eqn\excl{
|m^1-m^2|\leq m^3\leq {\rm min}\{m^1+m^2,\;r-m^1-m^2\}\,,
} because of the zero point energy factors as discussed in~\AldayRS.
This condition, when applied to \invtrin, 
changes the factors $\mmod{m^i-m^j-m^k}=0,\;r+m^i-m^j-m^k$ depending on whether 
$m^i-m^j-m^k$ is zero or not, and $\mmod{m^1+m^2+m^3}=0,\;m^1+m^2+m^3$
depending whether $m^1+m^2+m^3=r$ or not.

\

It is convenient to define a new operator $\hat {\cal H}_z$ related to 
${\cal H}_z$ by a conjugation,
\eqn\conj{
\hat {\cal H}_z=K^{-1}\, {\cal H}_z\, K\,,\qquad K=\frac{1}{\prod_{\vec{s}=\pm 1} \left((z^1)^{s^1}(z^2)^{s^2}(z^3)^{s^3}\,t\,q^{\mmod{-2\vec{m}\cdot \vec{s}}};q^r\right)}\,.
} This operator takes the following form
\eqn\basicopC{
\eqalign{
\hat {\cal H}_z\, F(\{m,\, z\})&=
 \cr
&\left[
\frac{1-t\,q^{r-2m}\,z^{2}}{1-q^{r-2m}\,z^2}\,F(\{m,\,q^{\frac{1}{2}r}\,z\})+
\frac{1-t\,q^{2m}\,z^{-2}}{1-q^{2m}\,z^{-2}}\,F(\{m,\,q^{-\frac{1}{2}r}\,z\})\right]+\cr
&\qquad\qquad+\frac{(1-t)(1-q^{-r})}{(1-q^{2m-r}\,z^{-2})(1-q^{-2m}\,z^{2})}\,F(\{m,\,q^{2m-\frac{1}{2}r}\,z^{-1}\})\,.
}
} 
As a matter of fact $\hat {\cal H}_z$ is a well-known object in mathematical literature,
as we will discuss in the next sub-section.
The eigenfunctions of this operator are neatly given in terms of Macdonald polynomials\foot{Note that the 
eigenfunction satisfy
$\phi^i_\ell(\{m,q^{k}\,z\})=\phi^i_\ell(\{m-k,\,z\})$.}
\eqn\macs{
\eqalign{
&\psi^1_\ell(\{m,z\})=P_\ell(q^{-m}\,z;q^r,\,t)\,,\cr
&\qquad\hat {\cal H}_z\,\psi^1_\ell(\{m,z\})=q^{-\frac{\ell}{2}\,r}(1+t\, q^{\ell\,r})\,\psi^1_\ell(\{m,z\})\,,\cr
&\psi^2_{\ell>0}(\{m,z\})=
S_\ell(q^{-m}\,z;q,\,t)\equiv 
(q^{-m} z-t\,q^{m}z^{-1})\,P_{\ell-1}(q^{-m}\,z;q^r,\,q^r\,t)\,,  \cr 
&\psi^2_{l=0}(\{m,z\})=0\ , \cr
&\qquad 
\hat {\cal H}_z\,\psi^2_\ell(\{m,z\})=q^{-\frac{\ell}{2}\,r}(1+t\,q^{\ell\,r})\,\psi^2_\ell(\{m,z\})\,,
}
}
where the polynomial $P_\ell(z;\,q, \, t)$ is the usual ({\it i.e.}
symmetric) Macdonald polynomial \MacdonaldBook.
For $A_1$ case this polynomial is symmetric, {\it i.e} 
$P_{\ell}(z;\, q, \, t)=P_{\ell}(z^{-1};\, q, \, t)$,
and is given by
\eqn\macdonaldsutwo{
P_\ell(z;\,q,\,t)=\frac{(q;q)_{\ell}}{(t;q)_{\ell}}
\,\sum_{i=0}^\ell
\frac{(t;q)_i}{(q;q)_i}
\frac{(t;q)_{\ell-i}}{(q;q)_{\ell-i}}\; z^{\ell-2i} \,.
} The  polynomial $S_\ell(z;q,\,t)$ is called the $t$-antisymmetric
Macdonald Polynomial \Marshall,
and as the name suggests is not a symmetric function.
The operators $\hat {\cal H}_{z}$ are self-adjoint under the natural measure of our problem: the measure 
with which we glue Riemann surfaces together, {\it i.e.} the vector multiplet measure,
\eqn\measMAcdo{
\Delta({\bf z})=\prod_{i<j}\frac{\theta(q^{m_j-m_i}\,z_i/z_j;q^r)}{\theta(t\,q^{m_j-m_i}\,z_i/z_j;q^r)}\,.
} Here we write the measure for general $A_{N-1}$ case
 and assume as before that for $i>j$ $m_i>m_j$. This is precisely the measure under which 
the non-symmetric Macdonald polynomials are orthogonal, see {\it e.g.}~\Marshall.

Note the degeneracy of the spectrum: the two eigenfunctions $\psi^i_\ell$ have the same
eigenvalue under the operator $\hat {\cal H}_z$ and they also are independent of $m$.
In particular the discussion up to this point implies that we can write the lens index of a generic theory of class ${\cal S}$ of type $A_1$ (genus $g$ and $s$ punctures) 
in the following form
\eqn\diagindex{
{\cal I}(\{m^i,z^i\})=
\sum_{\ell=0}^\infty\sum_{\gamma_i=1}^2 C^{(g)\;\gamma_1\cdots\gamma_s;m^1\cdots m^s}_\ell\,
\prod_{j=1}^s\phi^{\gamma_j}_\ell (\{m^j,\,z^j\})\,,\qquad \phi =K\,\psi\,.
} This structure was derived in the further Schur limit, $t=q^r$,
in~\AldayRS\ by explicitly studying the 
trinion theory.
We did not use all the constraints following from S-duality in writing~\diagindex. 
For the $r=1$ case one can fully exploit such constraints to completely fix the structure constants
$C^{(g)\;\gamma_1\cdots\gamma_s;m^1\cdots m^s}_\ell$~\GaiottoXA. We leave 
the problem of figuring out whether 
this is also the case for $r>1$ for future work.


\subsec{$A_{N-1}$ --- Cherednik Operators}

We can also evaluate the difference operators for the $A_{N-1}$ cases with $N>2$;
the explicit expression for the $A_2$ case can be found in appendix B.
Surprisingly, it turns out that the difference operators we obtain in the Macdonald limit are related to a
well studied object in mathematics, the {\it (double) affine Hecke algebra} ((D)AHA)
and especially the Cherednik operators \cherednik.

Let us thus make a brief interlude to define this mathematical structure.
To define AHA, one introduces the following operators acting on 
functions $f(\hat z_1, \cdots, \hat z_N)$ of $N$ variables
\eqn\AHA{
\eqalign{
&\sigma_i\,f(\cdots \,, \hat z_i,\,\hat z_{i+1},\cdots)=f(\cdots\,, \hat z_{i+1},\, \hat z_i,\cdots)\,,\cr
&\tau_i \, f(\hat z_1,\cdots,\hat z_N)=f(\hat z_1,\cdots,q\,\hat z_i,\cdots \hat z_N)\,,\cr
&T_i=t+\frac{t\;\hat z_i-\hat z_{i+1}}{\hat z_i-\hat z_{i+1}}(\sigma_i-1)\,,\qquad i=1,\cdots ,N-1\,,\cr
&T_0=t+\frac{q\;t\;\hat z_N-\hat z_{1}}{q\;\hat z_N-\hat z_{1}}(\sigma_N\,\tau_1\tau_N^{-1}-1)\,,\cr
&\omega=\sigma_{N-1}\,\sigma_{N-2}\cdots\sigma_2\,\sigma_1\,\tau_1\,.
}
} 
The operator $T_i$, sometimes called the Demazure-Lusztig operator, is a deformation of the permutation
$\sigma_i$ by  two parameters $q, t$.
It follows from this definition that the operators satisfy a set of nice relations,
\eqn\AHArel{\eqalign{
&(T_i-t)(T_i+1)=0\,,\qquad {\it i.e.}  \quad T_i^{-1}=t^{-1}-1+t^{-1}
\,,\cr
&T_i\,T_{i+1}\,T_i=T_{i+1}\,T_{i}\,T_{i+1}\,,\qquad \omega\,T_i=T_{i-1}\,\omega,\cr
&[T_i,\,T_j]=0\,,\qquad |i-j|\geq 2\,.
}
} 
These are the defining relations of the AHA.\foot{AHA
is defined by $T_i$'s and $\omega$ satisfying \AHA,
and equivalently can be defined in terms of $T_i$'s
and $Y_i$'s satisfying certain defining
relations. We can also define an operator $X_i$
which acts as a multiplication by $\hat z_i$,
and it turns out that $T_i$'s and $X_i$'s
also define another AHA. 
We can combine all of $T_i, X_i, Y_i$ containing two AHAs,
and this is known as the DAHA. See \cherednik\ for details.
}
Let us define the Cherednik operators $Y_i$ by
\eqn\chered{
\eqalign{
Y_i=t^{i-N}\; T_i \, T_{i+1}\, \cdots T_{N-1} \omega \, T^{-1}_1\,T^{-1}_2\,\cdots T^{-1}_{i-1}\, , \qquad i=1, \cdots, N \, .
}
} 
We can show from the relations \AHArel\ that
all these operators commute with each other \foot{
We also have $\prod_i Y_i=\omega^N$, where $\omega^N f(\hat{z}_1, \cdots, \hat{z}_N)=f(q\hat{z}_1, \cdots, q\hat{z}_N)$,
and hence only $N-1$ of the $Y_i$'s are in practice sufficient.
}
\eqn\Ycommute{
[Y_i,\, Y_j]=0 \, .
}

\

This mathematical structure is then related to the lens index in the following way.
After redefinition 
\eqn\redef{
\hat z_{N-i}=q^{-m_i+\frac1N \sum_{\ell=1}^Nm_\ell}\,z_i \ , 
}
and the identification $(t\to t,\, q^r\to q)$
 the operator of~\macresidue\  computing the residue at $a^*=t^{\frac12}q^{\frac{1}{N}}$
for $A_{N-1}$ quiver theories
is given by,
\eqn\symmOp{
K^{-1}\,{\cal H}\,K\sim  \sum_{\ell=1}^N Y_\ell \,,
} where
\eqn\Kfact{
 K=\prod_{i\neq j}^N\frac{1}{(t\, q^{\mmod{m_j-m_i}}\,z_i/z_j; q^r)}\,,
}
and $\sim$ in \symmOp\ represents the equality up to an overall
multiplicative factor which
depends only on $q$ and $t$.
We have checked this for $A_2$ case,
whose explicit difference operator can be found in appendix B.

\

The lens index in the Macdonald limit  of general $A_{N-1}$ quiver 
is then naturally given in terms of eigenfunctions of ${\cal H}$ which are 
$K$ times the non-symmetric Macdonald polynomials. 
Since the spectrum of non-symmetric Macdonald polynomials is degenerate 
the index will not be completely diagonal in this basis as we already discussed in the
 $A_1$ case.  We expect that the Cherednik operators will play a prominent role
if one will be interested in higher residues and in difference operators which compute
indices in presence of surface defects labeled by general representations of $A_{N-1}$.
We briefly discuss this issue in appendix C. 

\

\newsec{Large $r$ Limit}

Let us   comment on the lens index in the limit of large $r$.
In this limit the size of the Hopf fiber  of $S^3$ shrinks to zero length and we are left with $S^2$. One can think thus of the lens
index in $r\to\infty$ limit as the $S^2\times S^1$, {\it a.k.a.} the index, of the $3d$ theory 
with same matter content and the same global symmetry as the $4d$ theory. The lattice $(\bbZ_r)^{N-1}$ on which 
the index is defined becomes non-compact ($\bbZ^{N-1}$). 

For simplicity we focus on the $A_1$ case,
where the lattice is simply given by the line of integers $\bbZ$.
We define
\eqn\rescalings{
x=\sqrt{pq}\,, \qquad y=\sqrt{q/p}\,, \qquad
{\cal I}_m(b)={\cal J}_m(\left(p/q\right)^{m/2}\,b)\,,\qquad \left(p/q\right)^{m/2}\,b=\beta\,.
} 
Taking $r\to \infty$ the only residues are at $a^*=t^{\frac12}(pq)^{\frac{r}2}$.
The basic difference operator for $a^*=(p\,q\,t)^{\frac12}$ becomes
\eqn\redefO{
{\cal O}_{(p\,q\,t)^{\frac12}}\,\cdot\,{\cal J}=F^{m}_1\,{\cal J}_m(x\,\beta)+
F^{m}_2\,{\cal J}_m(x^{-1}\,\beta)+
G_1^{m}\,{\cal J}_{m+1}(\beta)+
G_2^{m}\,{\cal J}_{m-1}(\beta)\,,
} where  we have for $-\infty < m<\infty$ (recall \FpoleTwo)
\eqn\FpoleTwoLargeRred{
\eqalign{
&F^{m}_1=\left({\cal I}_V^{(m=0)}\right)^{-1}\,\frac{1-t^{\pm1}}{1-x^{\pm2}}\frac{(1-x^{2m}\frac{t}{x^2}\beta^{-2})(1-x^{2m}\frac{x^2}{t}\beta^{2})}{(1-x^{2m}\beta^{-2})(1-x^{2m}\beta^{2})}\,,\cr
&F^{m}_2=\left({\cal I}_V^{(m=0)}\right)^{-1}\,\frac{1-t^{\pm1}}{1-x^{\pm2}}\frac{(1-x^{2m}\frac{x^2}{t}\beta^{-2})(1-x^{2m}\frac{t}{x^2}\beta^{2})}{(1-x^{2m}\beta^{-2})(1-x^{2m}\beta^{2})}\,,\cr
&G_1^{m}=\left({\cal I}_V^{(m=0)}\right)^{-1}\,\frac{t}{x^2}\frac{1-t^{\pm1}}{1-x^{\pm2}}\frac{(1-x^{2m}\frac{x^2}{t}\beta^{-2})(1-x^{2m}
\frac{x^2}{t}\beta^{2})}{(1-x^{2m}\beta^{-2})(1-x^{2m}\beta^{2})}\,,\cr
&G_2^{m}=\left({\cal I}_V^{(m=0)}\right)^{-1}\,\frac{x^2}{t}\frac{1-t^{\pm1}}{1-x^{\pm2}}
\frac{(1-x^{2m}\frac{t}{x^2}\beta^{-2})(1-x^{2m}\frac{t}{x^2}\beta^{2})}{(1-x^{2m}\beta^{-2})(1-x^{2m}\beta^{2})}\,.
}
} 
Note that the dependence on $y$ drops out completely. This is to be
expected since the fugacity $y$ couples to the momentum along 
the Hopf fiber (recall \Idef) which shrinks in the large $r$ limit.

One can check the consistency of S-duality following similar discussion
of the previous section.
For example, the index of the trinion in the large $r$ limit is given by 
\eqn\trinionTD{\eqalign{
&{\cal J}_{H, \{m_i\}}(a,b,c)= \cr
&\quad \left(\frac{x^2}{t}\right)^{\alpha(m^1,m^2,m^3)}
\frac{
(t^{\frac12}\frac{x^2}t\,x^{|m^1+m^2-m^3|} (\frac{ab}{c})^{\pm1};x^2)
(t^{\frac12}\frac{x^2}t\,x^{|m^1+m^3-m^2|} (\frac{ac}{b})^{\pm1};x^2)}
{(t^{\frac12}\,x^{|m^1+m^2-m^3|} (\frac{ab}{c})^{\pm1};x^2)
(t^{\frac12}\,x^{|m^1+m^3-m^2|} (\frac{ac}{b})^{\pm1};x^2)}\cr
&\qquad \qquad\qquad  \times \frac{
(t^{\frac12}\frac{x^2}t\,x^{|m^3+m^2-m^1|} (\frac{cb}{a})^{\pm1};x^2)
(t^{\frac12}\frac{x^2}t\,x^{|m^1+m^2+m^3|} (abc)^{\pm1};x^2)}
{(t^{\frac12}\,x^{|m^3+m^2-m^1|} (\frac{cb}{a})^{\pm1};x^2)
(t^{\frac12}\,x^{|m^1+m^2+m^3|} (abc)^{\pm1};x^2)}\,,
}} 
where $\alpha(m^1, m^2, m^3)$ represents the zero-point contribution \zeropoint.\foot{
Note that the zero point contribution $\alpha(m^1,m^2,m^3)$ here includes also the holonomy $m_a$~\refs{\BeniniNC,\AldayRS}.
For example when $m_i$ satisfy triangle inequality and are positive $\alpha(m^1,m^2,m^3)=m^1+m^2+m^3$.
}
Acting on the trinion with the operator ${\cal O}_{(p\,q\,t)^{\frac12}}$  
on any one of the three $SU(2)$ flavor fugacities one obtains the same result,
as is expected from S-duality ({\it cf.} \propSduality).

One curious observation is that ${\cal O}_{(p\,q\,t)^{\frac12}}$ simplifies tremendously if 
we make the following Ansatz
\eqn\defJ{
{\cal J}_m(\beta)={\cal J}(x^{-m}\,\beta)\,,
} and then
\eqn\Jconseq{
{\cal J}_{m+1}(\beta)={\cal J}_m(x^{-1}\beta)\,,\qquad
{\cal J}_{m-1}(\beta)={\cal J}_m(x^{}\beta)\,.
} The difference operator~\redefO\ becomes
\eqn\macOP{
{\cal O}_{(p\,q\,t)^{\frac12}}\,\cdot\,{\cal J}=\frac{(1-t)^2(x^2+t)}{t\,(1-x^2)^2}\;\left[\frac{1-\frac{x^2}{t}\,\alpha^2}{1-\alpha^2}\;{\cal J}(x\,\alpha)+
\frac{1-\frac{x^2}{t}\,\alpha^{-2}}{1-\alpha^{-2}}\;{\cal J}(x^{-1}\,\alpha)\right]\,,
} where $\alpha=x^{-m}\,\beta$. This is precisely proportional to $A_1$ Macdonald operator $T_{\hat q,\,\hat t}$ with 
parameters $\hat t=\frac{x^2}{t}$ and $\hat q=x^2$.
Thus the operator ${\cal O}_{(p\,q\,t)^{\frac12}}$ can be thought of as yet another generalization
of the Macdonald operator.
 Note however that the index of the trinion
\trinionTD\ is not of the form \defJ\ and thus this Ansatz does not hold in our case.

\

\newsec{Final Remarks}

In this paper we have discussed an explicit procedure to obtain a set of difference
operators which act naturally on the lens space index of theories of class ${\cal S}$.
In particular finding the set of orthogonal eigenfunctions of these difference 
operators reduces the problem of fixing the lens index of theories of class ${\cal S}$ 
to a much simpler problem of finding a discrete set of structure constants.
It will be interesting to see whether the structure constants can be also fixed
only by assuming dualities.

We found surprisingly that
 the difference operators computing the resiudes in $U(1)$ fugacities of the lens index in the
Macdonald limit discussed in Sec.\ 4 can be nicely written in terms of the Cherednik operators
appearing in (D)AHA. It would be interesting to see if the (D)AHA for other root systems are
relevant for the study of lens indices for Gaiotto theories for $D_N$
 and $E_{6,7,8}$ ({\it cf.} \MekareeyaTN, \LemosPH).

(D)AHA has been discussed in a number of different contexts in 
mathematical physics. In particular it
has recently been used in the construction of knot invariants
(see \CherednikNR\ and subsequent works). For torus knots using (D)AHA
these papers  give knot invariants identical to those coming from the
refined Chern-Simons theory of \AganagicSG.
Since the latter appears to be closely related to the superconformal index (see
e.g.\ \GaddeUV ), 
the appearance of (D)AHA in the two different contexts is probably not a coincidence.
It would be interesting to explore this point further.

The structure of our lens index, with full fugacities $p, q, t$,
suggests that there is even richer mathematical structure when we incorporate the 
parameter $p$. This should be associated with some {\it elliptic} generalization of DAHA.
The situation is more complicated in this case since there are three different types of 
basic difference operators ${\cal O}_{t^{\frac{1}{2}}(pq)^{1/N}}, {\cal O}_{t^{\frac{1}{2}}p^{r/N}}, {\cal O}_{t^{\frac{1}{2}}q^{r/N}}$. These operators can be viewed as a matrix valued generalization of the 
``hamiltonians'' of the 
Ruijsenaars-Schneider integrable models.

Finally, it would be interesting to see if similar techniques could be applied to 
4d ${\cal N}=1$ theories, such as the theories in \BeniniMZ\ and 
those in \XieMR, \FrancoMM. The lens space indices of the latter
theories
will be discussed in \Yamazaki\ in connection with integrable models.

\bigskip
\noindent{\bf Acknowledgments:} 

\

We would like to thank 
M.~Noumi and B.~Willett for very useful and stimulating discussions and correspondence.
We would like to thank the organizers of the workshop ``Geometric Correspondences of Gauge Theories'', 
(September 2012, SISSA) for providing a stimulating environment, during which this project has 
been initiated. SSR gratefully acknowledges support from the Martin~A.~Chooljian and Helen Chooljian membership
 at the Institute for Advanced Study. The research of SSR was also partially supported by
NSF grant number PHY-0969448. The research of MY was supported by World Premier International
Research Center Initiative (WPI Initiative), MEXT, Japan.


\appendix{A}{A Technical Proof}

In this appendix we show that the rescaled index \rescaledtrinion\ 
is  well-defined  in the Macdonald limit $p\to 0$,
{\it i.e.}, the rescaled index has only non-negative powers of $p$
and the only remaining contribution is \condmac.

The power of the zero-point contribution of \rescaledtrinion\ is given by

\eqn\sumf{
\frac{2}{4r } \left[
\sum_{i,j} f(m_i+\widetilde{m}_j) -\sum_{i<j} f(m_i-m_j)-\sum_{i<j} f(\widetilde m_i-\widetilde m_j)
\right] \ ,
}
with $f(x):= \mmod{x} (r-\mmod{x})$.
Note that we have $f(0)=f(r)=0$ and $f(x)=f(-x)$.
Since the expression \sumf\ is invariant under the permutation of $m_i$'s and also of $\widetilde m_i$'s
we can assume without generality that
\eqn\ineq{
0\le m_N\le m_{N-1} \le \cdots \le m_2\le x_1< r  \ , \qquad
1\le \bar m_N\le \bar m_{N-1} \le \cdots \le \bar m_2\le \bar m_1< r  \ ,
}
where we defined $\bar m_i:=r-\widetilde m_i$. The expression \sumf\ then becomes
\eqn\sumftwo{
\frac{2}{4r } \left[
\sum_{i,j} g(|m_i-\bar{m}_j|) -\sum_{i<j} g(m_i-m_j)-\sum_{i<j} g(\bar m_i-\bar m_j)
\right] \ ,
}
with $g(x):=x(r-x)$.
The expression inside the bracket of \sumftwo\ is non-negative:
\eqn\sumfthree{
\eqalign{
&r \left( 
\sum_{i,j} |m_i-\bar m_j| -\sum_{i<j} (m_i-m_j+\bar m_i-\bar m_j)
\right)  \cr
& \qquad+\left( 
\sum_{i,j} (m_i-\bar m_j)^2 -\sum_{i<j} (m_i-m_j)^2-\sum_{i<j} (\bar m_i-\bar m_j)^2
\right) \cr
& =r\left( \sum_{i< j}\left( |m_i-\bar m_j| -(m_i-\bar m_j) 
+ |m_j-\bar m_i| +(m_j-\bar m_i) \right)  
+\sum_{i}  |m_i-\bar m_i| 
\right)  \cr
& \qquad \qquad +\left( 
\sum_{i} (m_i-\bar m_i) \right)^2 \ge 0 \ ,
} 
}
where the equality holds only when $m_i=\bar m_i$ for all $i$. This gives \condmac, 
after lifting the condition \ineq.

\appendix{B}{$A_2$ Macdonald Limit}

In this appendix we quote the difference operator 
computing the basic residue in the Macdonal limit for the
$A_2$ case.
The basic difference operator, associated with the pole $a^*=t^{\frac{1}{2}}\,q^{\frac{r}{3}}$, 
can be written in terms of 
seven different functions $F_I,\, G_{(I J)}$ and
$ G_{(1\,3\,2)}$ in ~\macresidue:
\eqn\Hnew{
\eqalign{
&K^{-1}\,{\cal H}\,K\, f(z_1,z_2,z_3)\sim
\cr
&\qquad \widetilde F_1\,f(q^{-\frac{2}{3}}z_1,q^{\frac{1}3}z_2,q^{\frac{1}3}z_3) +\widetilde F_2\,f(q^{\frac{1}3}z_1,q^{-\frac{2}{3}}z_2,q^{\frac{1}3}z_3)+
\widetilde F_3\,f(q^{\frac{1}3}z_1,q^{\frac{1}3}z_2,q^{-\frac{2}{3}}z_3)\cr
&
\qquad \qquad +
\widetilde G_{(1\,2)}\,f(q^{\frac{1}{3}+m_2-m_1}z_2,q^{-\frac{2}3+m_1-m_3}z_1,q^{\frac{1}3}z_3)  \cr
& \qquad \qquad +\widetilde G_{(1\,3)}\,f(q^{\frac{1}3+m_3-m_1}z_3,q^{\frac{1}{3}}z_2,q^{-\frac{2}3+m_1-m_3}z_1)\cr
&\qquad\qquad 
+ \widetilde
 G_{(2\,3)}\,f(q^{\frac{1}3}z_1,q^{\frac{1}3+m_3-m_2}z_3,q^{-\frac{2}{3}+m_2-m_3}z_2)
\cr
&\qquad \qquad  +\widetilde G_{(1\,2\,3)}\,f(q^{\frac{1}3+m_2-m_1}z_2,q^{\frac{1}3+m_3-m_2}z_3,q^{-\frac{2}{3}+m_1-m_3}z_1)
\, .
}
}
Assuming for concreteness that $r> m_1>m_2>m_3>0$ and $m_1+m_2+m_3=r$, we find
\eqn\Fs{
\eqalign{
&\widetilde F_1\qquad:\qquad \frac{1-t}{1-q^{-r}}\,\frac{1-t\, q^r\,q^{m_2-m_1}\frac{z_1}{z_2}}{1-q^r\,q^{m_2-m_1}\frac{z_1}{z_2}}\,
\frac{1-t\,q^r\,q^{m_3-m_1}\frac{z_1}{z_3}}{1-q^r\,q^{m_3-m_1}\frac{z_1}{z_3}}\,,\cr
&\widetilde F_2\qquad:\qquad 
\frac{1-t}{1-q^{-r}}\,\frac{1-t\,q^{m_1-m_2}\frac{z_2}{z_1}}{1-q^{m_1-m_2}\frac{z_2}{z_1}}\,
\frac{1-t\,q^r\,q^{m_3-m_2}\frac{z_2}{z_3}}{1-q^r\,q^{m_3-m_2}\frac{z_2}{z_3}}\,,\cr
&\widetilde F_3\qquad:\qquad 
\frac{1-t}{1-q^{-r}}\,\frac{1-t\,q^{m_1-m_3}\frac{z_3}{z_1}}{1-q^{m_1-m_3}\frac{z_3}{z_1}}\,
\frac{1-t\,q^{m_2-m_3}\frac{z_3}{z_2}}{1-q^{m_2-m_3}\frac{z_3}{z_2}}\,,\cr
}
}
\eqn\Gs{
\eqalign{
&\widetilde G_{(1\,2)}\qquad:\qquad (1-t)^2\,\frac{1-t\, q^r\,q^{m_3-m_1}\frac{z_1}{z_3}}
{(1-q^{m_2-m_1}\frac{z_1}{z_2})(1-q^r q^{m_3-m_1}\frac{z_1}{z_3})
(1-q^{-r}q^{m_1-m_2}\frac{z_2}{z_1})}\,,\cr
&\widetilde G_{(1\,3)}\qquad:\qquad (1-t)^2\,\frac{1-t\,q^{m_2-m_3}\frac{z_3}{z_2}}
{(1-q^{m_2-m_3}\frac{z_3}{z_2})(1-q^{-r} q^{m_1-m_3}\frac{z_3}{z_1})
(1-q^{m_3-m_1}\frac{z_1}{z_3})}\,,\cr
&\widetilde G_{(2\,3)}\qquad:\qquad (1-t)^2\,\frac{1-t\,q^{m_1-m_2}\frac{z_2}{z_1}}
{(1-q^{m_1-m_2}\frac{z_2}{z_1})(1-q^{-r} q^{m_2-m_3}\frac{z_3}{z_2})
(1-q^{m_3-m_2}\frac{z_2}{z_3})}\,,\cr
&\widetilde G_{(1\,2\,3)}\qquad:\qquad (1-t)^3\,\frac{1}
{(1-q^{m_2-m_1}\frac{z_1}{z_2})(1-q^{-r} q^{m_1-m_3}\frac{z_3}{z_1})
(1-q^{m_3-m_2}\frac{z_2}{z_3})}\,.
}
} 
As explained in the main text, we can explicitly verify that 
this difference operator is conjugate to the sum of 
$A_2$ Cherednik operators $Y_{1,2,3}$:
\eqn\HtoY{
K^{-1}\,{\cal H}\,K  \sim Y_1+Y_2+Y_3\, ,
}
where we used the parameter identification \redef, $K$ is given in \Kfact, and $Y_i$ can be computed from the definition \AHA, \chered.
Note that we do not have a symmetry interchanging the three indices; for
example $\tilde F_{1}$ is different from $\tilde F_{2}$ even after the exchange of 
$z_i$'s and $m_i$'s.
In the index computation this follows from a particular ordering of the 
holonomies $m_i$ and in the affine Hecke algebra from the non-symmetric definition
of $Y_i$~\chered. 

\appendix{C}{More Comments on Difference Operators in the Macdonald Limit}

The discussion of this paper can be generalized to include difference operators associated
to general irreducible representations of $A_{N-1}$.
In the $r=1$ case it was argued in \GaiottoXA\ that the difference operators computing residues 
at $a^*=t^{\frac12}q^{\frac{n}N}$ for $n=1,\cdots,N-1$ correspond to introducing 
certain surface defects 
to the index computation which are associated to the $n$th symmetric representation of $A_{N-1}$.
One can then discuss  difference operators corresponding to introducing surface operators assocaited 
to more general representations. 
Such a generalization was recently elaborated upon in \AldayKDA\ for the 
ordinary superconformal index $r=1$ in the Schur limit $p=0, q=t$. Here we will comment on the lens
version of this generalization.

The finite irreducible representations of $A_{N-1}$ are in one to one corerspondence with 
Young diagrams. For a surface
defect represented by a  Young diagram ({\it a.k.a} partition) $\lambda$, we propose that the associated 
difference operator ${\cal H}_{\lambda}$ satisfies
\eqn\Yconj{
K^{-1}\, {\cal H}_{\lambda} \, K\sim s_{\lambda}(Y) \ ,
}
where $K$ is the same operator \Kfact\
defined previously and $s_{\lambda}$ is the Schur polynomial associated
with $\lambda$
and we have again neglected the overall constant multiplicative factor.
Note that the order of $Y$'s does not matter in $s_{\lambda}(Y)$ since $Y$'s commute with each other \Ycommute.

The proposal \Yconj\ passes several non-trivial tests.
First for a fundamental representation we have
$s_{\lambda}(Y)=\sum_{i=1}^N Y_i$, so \Yconj\ reduces to 
\symmOp.
Second, we learn immediately from \Yconj\ and \Ycommute\
that the operators ${\cal H}_{\lambda}$'s commute:
\eqn\Hcommute{
[{\cal H}_{\lambda}, {\cal H}_{\mu}]=0\ .
}
This is consistent with the expectations from the S-duality, see the
similar discussion in Sec.\ 3.
Third, \Yconj\ is consistent with the 
decomposition of the tensor product of representations,
$R_{\lambda}\otimes R_{\mu}=\sum_{\nu} N_{\lambda, \mu}^{\nu} R_{\nu}$,
where $N_{\lambda, \mu}^{\nu}$ is the Littlewood-Richardson coefficient.
The Schur functions, being the character of irreducible representations,
 satisfy the corresponding statement
\eqn\schurdecomp{
s_{\lambda}(Y) s_{\mu}(Y)= \sum_{\nu} N_{\lambda, \mu}^{\nu} s_{\nu}(Y) \ ,
}
This represents operator product expansions 
of the surface operators, and is
similar to the decomposition of degenerate fields in the Liouville theory.

Finally, \Yconj \ is consistent with the known results for the $r=1$ index.
When restricted to the $r=1$ case the operators act on the symmetric
functions, since there no non-trivial discrete Wilson lines
and hence the Weyl group of the gauge group is always unbroken.
It is known that \symmOp\ acting on symmetric functions coincide
with the corresponding Macdonald operators. More formally, there
is a $\bbQ(q,t)$-algebra isomorphism \Noumi
\eqn\YM{
{\bbQ}(q,t)[Y_1^{\pm 1}, \cdots , Y_N^{\pm 1}]^{{S}_N}   \simeq \bbQ(q,t)[D_1, \cdots, D_{N-1}, D_N^{\pm 1} ] \, ,
}
where $S_N$ is the Weyl group of $A_{N-1}$, {\it i.e.}, 
we have symmetric polynomials of $Y_i$ on the left hand side.
On the right we have $k$-th
Macdonald operators 
\eqn\MacOp{
D_k:=t^{-r(n-r)} 
\sum_{I\subset \{ 1, \cdots, N \}: \, |   I |=r}
\prod_{i \in I, j\not \in I} 
\frac{t z_i-z_j}{z_i-z_j} \prod_{i\in I} \tau_i   \,, \qquad k=1, \cdots,
N\, ,
}
where $\tau_i$ are defined to be the $q$-multiplication of one of the
arguments
as in \AHA. 
When $\lambda$ is the $k$-th antisymmetric representation
we have $s_{\lambda}(Y)=\sum_{i=1}^N Y_i^k$,
and the image of the map under the 
isomorphism \YM \ is given by $D_k$ (up to powers of $t$),
as can be checked by explicit computations \KirillovNoumi.
The simplest case is the $A_1$ theory, where we can easily show that
the operator \basicopC\ acting on a symmetric function coincides
with the operator $D_1$.

\

The eigenfunctions of the operator \Yconj\ are known as non-symmetric Macdonald polynomials,
generalizing our explicit  observation for the $A_1$ case.
Given a composition $\eta=(\eta_1, \eta_2, \cdots)$ of non-negative integers, 
there is an associated non-symmetric Macdonald polynomial $E_{\eta}(z;q,t)$.
This is the simultaneous eigenfunction of the Cherednik operators $Y_i$
(see \MacdonaldHecke, \cherednik, \Marshall ):
\eqn\Yeigen{
Y_i\, E_{\eta}(x;q,t)=q^{\eta_i} t^{-l_{\eta}'(i)}
E_{\eta}(x;q,t) \ , \qquad i=1, \cdots, N \, ,
}
where we defined the leg co-length $l_{\eta}'(i)$ by
\eqn\lprime{
l_{\eta}'(i):=
\#\{ k\, |\, k<i, \eta_k \ge \eta_i \}
+
\#\{ k\, |\, k>i, \eta_k > \eta_i \} 
 \ .
} 
For each composition $\eta$ we can permute the elements
such that the resulting expression is a partition. We denote this
partition by $\eta^+$.
It follows from \Yeigen\ that $E_{\eta}(z;q,t)$
is the eigenfunction of $s_{\lambda}(Y)$ with eigenvalue 
$s_{\lambda}(q^{\eta_i} t^{-l'_{\eta}(i)} ) $, which is 
determined solely in term of the associated partition $\eta^+$.
In other words $E_{\eta}(z;q,t)$ have the same eigenvalues under the operator
$s_{\lambda}(Y)$ as long as the associated partitions $\eta^+$ are the same.
This means that we have degeneracies of the spectrum,
as we have already observed for the $A_1$ case.\foot{The $A_N$ 
analogues of symmetric and $t$-antisymmetric polynomials are 
given by
\eqn\tantisymm{
P_{\eta^+}(z;q,t)=\frac{1}{\gamma^+_{\eta}(q,t)}U^+ E_{\eta}(z;q,t)\,, \qquad
S_{\eta^+}(z;q,t)=\frac{1}{\gamma^-_{\eta}(q,t)}U^- E_{\eta}(z;q,t) \,,
}
where $\gamma_{\eta}^{\pm}(q,t)$ are normalization constants
and $U^{\pm}$ are the operators representing $t$-symmetrization and $t$-antisymmetrization:
\eqn\symmetrization{
U^+=\sum_{\sigma\in {S}_N} T_{\sigma}\, \qquad
U^-=\sum_{\sigma\in {S}_N} (-t)^{-l(\sigma)} T_{\sigma} \,
}
where $l(\sigma)$ is the sign of the permutation ($l(\sigma):=\#\{(i,j) | i<j, \sigma_i>\sigma_j \}$)
and $T_{\sigma}$ is the product of $T_{i}$'s
when $\sigma$ is decomposed into a product of adjacent transpositions:
$T_{\sigma}=T_{i_1} T_{i_2} \cdots $ when $s=s_{i_1} s_{i_2} \cdots$.
}

In mathematics many of the properties of the 
ordinary Macdonald polynomials $P_{\lambda}(x;q,t)$ are more transparent when we consider more general
non-symmetric Macdonald polynomials $E_{\eta}(x;q,t)$.
This is mirrored in our physics discussion,
where the properties of the superconformal index and the
difference operators acting on them are more transparent in the 
general lens indices.
For example, the decomposition of the product of difference operators
in \schurdecomp\ follows trivially from the commutativity of 
Cherednik operators, whereas similar statements 
in terms of
Macdonald operators $D_k$ and symmetric functions
are more involved.

\listrefs
\end